\documentclass[journal]{IEEEtran}
\IEEEoverridecommandlockouts
\usepackage{cite}
\usepackage{amsmath,amssymb,amsfonts, amsmath, dsfont}
\usepackage{graphicx}
\usepackage{textcomp}
\usepackage{xcolor}
\usepackage{multirow}
\usepackage{subcaption}
\usepackage{cite}
\usepackage{textcomp}
\usepackage{dirtytalk}

\usepackage{algorithm,algorithmic}

\usepackage{tikz}
\usepackage{xcolor, colortbl}
\usetikzlibrary{decorations.pathreplacing}
\usetikzlibrary{arrows,shapes,positioning}
\usetikzlibrary{patterns}
\usepackage{pgfplots}
\pgfplotsset{compat=newest}
\pgfplotsset{plot coordinates/math parser=false}
\pgfkeys{/pgf/number format/.cd,1000 sep={\,}}
\usetikzlibrary{plotmarks}

\newlength\matlabfigurewidth
\newcolumntype{L}[1]{>{\raggedright\let\newline\\\arraybackslash\hspace{0pt}}m{#1}}
\newcolumntype{C}[1]{>{\centering\let\newline\\\arraybackslash\hspace{0pt}}m{#1}}
\newcolumntype{R}[1]{>{\raggedleft\let\newline\\\arraybackslash\hspace{0pt}}m{#1}}

\begin{document}

\title{Optimal Planning of Single-Port and Multi-Port Charging Stations for Electric Vehicles in Medium Voltage Distribution Networks}

%
%

\author{Biswarup Mukherjee,~\IEEEmembership{Member,~IEEE,}
        and~Fabrizio~Sossan,~\IEEEmembership{Member,~IEEE}
\thanks{Both authors are with the Centre for processes, renewable energies and energy systems (PERSEE) of MINES ParisTech - PSL University, France. E-mail: {biswarup.mukherjee, fabrizio.sossan}@mines-paristech.fr}
\thanks{Project supported by the joint programming initiative ERA-Net Smart Energy systems’ focus initiative Integrated, Regional Energy Systems and European Union’s Horizon 2020 research and innovation programme under grant agreement No 775970, in the context of the EVA project.}
}


\maketitle

\begin{abstract}
This paper describes a method based on mixed-integer linear programming to cost-optimally locate and size chargers for electric vehicles (EVs) in distribution grids as a function of the driving demand. The problem accounts for the notion of single-port chargers (SPCs), where a charger can interface one EV maximum, and multi-port chargers (MPCs), where the same charger can interface multiple EVs. The advantage of MPCs is twofold. First, multiple ports allow arbitraging the charging among multiple vehicles without requiring the drivers to plug and unplug EVs. Second, the charger's power electronics is not sized for the total number of charging ports, enabling cost savings when the grid constraints are bottleneck of the problem. The proposed method can account for different charger typologies, such as slow and fast chargers, and model the drivers' flexibility of plugging and unplugging their EVs. Simulation results from a synthetic case study show that implementing MPCs is beneficial over both SPCs and drivers' flexibility in terms of total investments required for the charging infrastructure.

\end{abstract}

\begin{IEEEkeywords}
EVs; Charging stations; Distribution networks; 
\end{IEEEkeywords}

\section{Introduction}

The massive adoption of electric vehicles (EVs) will play a central role in decarbonizing road transportation \cite{EUCEF, 1irena_smartcharging, metais:hal-03127266, EV_dataPlot}.

Recharging EVs requires to develop an extended and pervasive charging infrastructure. Reference \cite{USAInvestment} estimates that, between 2019-2025, more than 2 billion Dollars will be necessary to improve the public and residential charging infrastructure across major U.S. metropolitan areas, whereas, in France, 2 billion Euros will be required to achieve the target of 7 million deployed public and private charger by 2030 \cite{metais:hal-03127266,MinistryReport}. In addition to these investments, others will be necessary to adapt the electrical grid infrastructure, in particular distribution grids. Indeed, it is well known that the connection of many chargers in distribution grids might determine congestions at the level of substation transformer and lines, and violations of statutory voltage limits (e.g., \cite{6674071, en12244717}). This is because distribution grids were designed to host prescribed amounts of demand and with predefined voltage gradients along the feeders, which are typically violated when massively recharging EVs. The large investments required to both install suitable charging infrastructure for EVs and upgrade existing distribution networks motivate the need to research formal methods to locate and size EV chargers effectively accounting for realistic driving demand patterns, technical limits of existing distribution grids, and cost of the chargers.

In this context, this paper proposes a method to locate and size EV chargers accounting for the driving demand and the constraints of existing distribution grids. The method can model different types of chargers, such as fast and slow chargers, as well as single-port chargers (SPCs) and multi-port chargers (MPCs). In addition, it can model the availability of the drivers to plug and unplug EVs to and from public charging stations, an element that can significantly impact the utilization of the charging columns, and possibly the number of chargers to deploy. The method is thought for an integrated distribution system operator (DSO)/urban planner, who wishes to design, or get insights on, a cost-optimal charging infrastructure while accounting for the technical limitations of the grid, different types of chargers, and drivers behaviors.

The problem of planning the EVs recharging infrastructure is not new and has been extensively investigated in the literature, although not in the terms proposed in this paper.
For example, a multi-objective planning model for the layout of electric vehicle charging station is proposed in \cite{HengsongWang}, without considering, however, distribution grid's operational constraints.
The work in \cite{6398568} proposes joint planning of EVs charging stations and distribution capacity expansion, without modeling, however, MPCs and drivers flexibility.
Authors of \cite{6362255} proposed a method for the cost-optimal planning of EV charging stations in a distribution grid considering grid constraints; however, this work did not consider MPCs that, as shown in this paper, can achieve significant cost savings. Methods for optimal planning of charging stations were also developed in \cite{6879337,7368203}, without however including grid constraints. 
The work in \cite{8000654} proposes a planning method to design  multiple-charger multiple-port charging systems for EVs that features the capability of sharing a limited number of chargers to more EVs. However, this method extends to a parking slot and not to whole distribution grid.
In \cite{en12132595}, and similarly in \cite{XIANG2016647}, both distribution network and traffic flows were used to identify appropriate nodes to locate and size the EV charging stations. This work uses genetic algorithm to solve nonconvex AC load flows, a formulation which could not scale well to a large number of EVs, and do not consider voltage and line ampacities constraints, only the rated power of the nodes. The work in \cite{8366991} proposes a data-driven approach for identifying driving demand and, based on this information, advises system planners on suitable locations for the charging infrastructure without considering, however, grid constraints.
More recently, in \cite{9265279, 9076708}, a two-stage optimization framework was proposed, in combination with an efficient resolution method, to co-optimize the charging infrastructure in combination with the operations of the power grid and gas network. However, the work does not specifically address multi-port chargers and drivers' flexibility.  In the light of the current state-of-the-art, the contributions of this paper are as follows: i), a planning method to locate and size chargers of EVs in distribution grids accounting accounting for grid constraints and multiple charger typologies, including slow chargers, fast chargers, SPCs, and MPCs; ii), a dedicated set of constraints to model the flexibility of the drivers in plugging and unplugging their vehicles into and from charging station; and, iii), a mixed-integer linear formulation embedding a linearized grid model that can be solved with off-the-shelf optimization libraries.

The paper is organized as follows. Section II describes the problem statement, input quantities, and model assumptions. Section III describes the models adopted in the planning problem and its formulation. Section IV describes how drivers' flexibility is modeled. Section V describes the synthetic case study adopted to test the models. Section VI presents results and discussions. Finally, Section VII concludes the paper.

\section{Problem Statement}
\begin{figure}
    \centering
    \includegraphics[width=0.8\columnwidth]{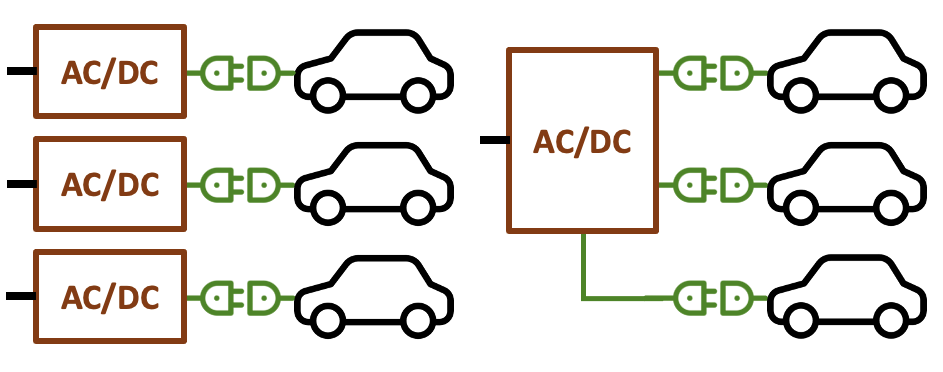}
    \caption{Single-port chargers (SPCs) on the left, and multi-port chargers (MPCs) on the right.}
    \label{fig:chargers}
\end{figure}

The objective of the problem is to identify the location and rating (i.e., fast and slow charging) of EV chargers in a distribution grid to satisfy the charging demand of a given population of vehicles while attaining minimum capital investment costs and respecting distribution grid's constraints. Moreover, we also want to model the operations of both conventional single-port chargers (SPCs) and multi-port chargers (MPCs), with the ultimate objective of evaluating the cost savings of one or the other configuration. The distinction between SPCs and MPCs is that, while SPCs has a plug for each charging column, MPCs have a centralized power conversion stage and multiple ports (Fig.~\ref{fig:chargers}). From a technological perspective, the differentiating factor between MPCs and SPCs is that MPCs, i), can have smaller power electronics ratings than SPCs for the same number of plugs and, ii), enables arbitraging the charge of the locally connected vehicles without requiring drivers to plug and unplug their vehicles.
Although the proposed methodology is general and can be adapted to model arbitrary power rating of the charging stations, we specifically consider two charger ratings for SPCs, i.e., fast and slow chargers, with kVA rating denoted by $\bar F$ and $\bar S$ (where, $\bar F > \bar S$), respectively. For MPCs, the charger rating is assumed to be a multiple of $\bar S$ or $\bar F$.

\subsection{Input information and notation}
The formulation uses two main sets of input information, one for the EVs and one for the power grid. For the EVs, their parking location and battery discharging power (depending on the driving demand) over time is assumed given (e.g., estimated from urban mobility data or statistics). An example of this is in the Results section of this paper. For the power grid, the grid topology, lines characteristics, and relevant nodal injections due to demand and distributed generation are assumed known, e.g., from measurements, state-estimation procedures, or statistics. This last set of information is necessary to model the operational constraints of the distribution grid so as to produce chargers deployment plans that are respectful of grid constraints.

The parking locations of the EVs over time is encoded in the following input binary variables:
\begin{align}
    p_{nvt} =
    \begin{cases}
    1, & \text{if EV }v\text{ is connected to node } n \text{ at time } t\\
    0, & \text{otherwise}\\
    \end{cases} \label{eq:parking}
\end{align}
where $n=1,\dots,N$ denote the node of the distribution grids, $t=1,\dots,T$ the time interval, and $v=1,\dots,V$ the index of the EVs. For more compact expressions, and with abuse of notation, we denote with subscripts $nvt$ quantities for node $n$, vehicle $v$ and at time interval $t$, and not the product among these indexes; similarly for other subscripts. It is illustrative to mention that a vehicle can be parked at one node only at a time, thus the following holds:
\begin{align}
 \sum_{n=1}^N p_{nvt} \le 1 && \text{for all $v$ and $t$}.
\end{align}

\subsection{Modeling assumptions}
We highlight here the general assumptions adopted for this study.  Specific assumptions related to certain modeling choices are remarked throughout the paper when used. It is considered the perspective of an integrated grid operator/urban planner wishing to attain minimum capital investments for the charging infrastructure. Operational costs (e.g., minimizing the cost of electricity) and operational strategies (e.g., optimal price-making strategies) are not considered, although the proposed methods could provide useful insights in that direction. Chargers are assumed to operate on/off. In other words, the recharging power cannot be modulated in intensity; however, it can be modulated in time, thus, from a grid perspective, achieving power intensity modulation at the aggregated level. In this respect, smart charging (in the sense of modulating the charging power for the grid benefit) is considered in the problem. Reactive power support from the chargers and V2G are not considered because, although they can help with grid congestions, their contribution is typically small (e.g., \cite{ISGT2021V2G}) and not expected to impact the planning results significantly.

\section{Methods}
As explained in this section, the problem of planning the EV charging infrastructure is formulated as a constrained economic optimization program. Its objective is to minimize the capital investment of the chargers while meeting the charging demand and respecting grid operational constraints. This section is structured as follows. First, the models for the plugged and charging states of the EVs is presented. Then, based on these models, it is explained how the needs for chargers are identified, along with their costs. Based on the charging states of the EVs and charger typologies, the nodal injections are formulated and used in a linearized load flow to approximate grid constraints. Finally, all these models are joined together in the planning problem. The next section will discuss which other constraints are needed in the planning problem to model drivers flexibility.

\subsection{State and charging power of the electric vehicles}
\subsubsection{Modeling connection and charging state}
For each vehicle $v$ and time interval $t$, the binary variables $f^{\text{plugged}}_{vt}, s^{\text{plugged}}_{vt}$ are defined to denote whether the vehicle is connected to a fast and a slow charger, respectively.  As an EV cannot be connected to a fast and slow charger simultaneously, it holds that $f^\text{plugged}_{vt}$ and $s^\text{plugged}_{vt}$ cannot be active at the same time; moreover, as an EV can be plugged only when parked, $f^\text{plugged}_{vt}$ and $s^\text{plugged}_{vt}$ can be active only if at least one $p_{nvt}$ among all nodes is active. These two requirements can be formalized as the following constraint:
\begin{align}
    f^\text{plugged}_{vt} + s^\text{plugged}_{vt} \le \sum_{n=1}^N p_{nvt} && \text{$\forall t$ and $v$} \label{eq:pluggedinonlyifparked}.
\end{align}

Two additional binary variables per vehicle and time interval, denoted by $f^{\text{charge}}_{vt}, s^{\text{charge}}_{vt}$, indicate whether a vehicle is charging from a fast or a slow charger. As EVs can charge only when plugged, it holds that:
\begin{subequations}
\begin{align}
    f^{\text{charge}}_{vt}  \le f^{\text{plugged}}_{vt} && \text{$\forall t$ and $v$}\\
    s^{\text{charge}}_{vt}  \le s^{\text{plugged}}_{vt} && \text{$\forall t$ and $v$}
\end{align}\label{eq:chargeonlyifparked}\end{subequations}

Quantities $f^\text{plugged}_{vt}, s^\text{plugged}_{vt}, f^{\text{charge}}_{vt}, s^{\text{charge}}_{vt}$ are variables of the optimization problem; based on these variables, the charging power of the EVs, as well as the needs for fast and slow chargers are determined as explained next.

\subsubsection{Charging power}
With the above definitions in place, the charging power of a vehicle $v$ and time $t$ is:
\begin{subequations}\label{eq:ev_charging_power}
\begin{align}
& p^{\text{EV+}}_{vt} = f^{\text{charge}}_{vt} \cdot \bar{F} \cdot \text{cos}\phi_F + s^{\text{charge}}_{vt} \cdot \bar{S} \cdot \text{cos}\phi_S \label{eq:ev_charging_power_a},
\end{align}
where input parameters $\bar F$ and $\text{cos}\phi_F$ are the kVA rating and power factor of the fast charger, respectively, and similarly for $\bar S$ and $\text{cos}\phi_S$. The reactive power associated to this charging demand is:
\begin{align}
& q^{\text{EV+}}_{vt} = f^{\text{charge}}_{vt} \cdot \bar{F} \cdot \text{sin}\phi_F + s^{\text{charge}}_{vt} \cdot \bar{S} \cdot \text{sin}\phi_S.
\end{align}
\end{subequations}

\subsubsection{Vehicles' state of charge}
The evolution of the vehicles' SOC over time is now modelled. This depends on the charging and discharging power. The charging power is as in \eqref{eq:ev_charging_power_a} and is determined by the planning problem. The discharging power is, instead, an input of the problem, as discussed hereafter. The SOC of vehicle $v$ at time $t$ is modeled as:
\begin{align}
\text{SOC}_{v}(t) =  \text{SOC}_{v}(0) + \frac{T_s}{E_v} \sum_{\tau=0}^{t-1} \left( \eta \cdot p^{\text{EV+}}_{v\tau}-  p^{\text{EV}-}_{v\tau} \right),  \label{eq:SOC model}
\end{align}
where $\text{SOC}_{v}(0)$ is the initial SOC (a problem decision variable, as it will be discussed later), $p^{\text{EV-}}_{vt}$ the discharging power in kW, $T_s$ the sampling time in hours, $E_v$ the nominal energy capacity of the EV's battery (in kWh), and $\eta$ is the charging efficiency. Model \eqref{eq:SOC model} is linear in the power; it is commonly adopted in the literature (e.g. \cite{JuanVanRoyThesis}) and assumes constant battery's voltage and efficiency. These assumptions, which trade-off accuracy for increased model tractability, can be considered acceptable in a planning problem with sparse temporal resolution (e.g., 1 hour). The vehicles' SOCs should be within a feasible range (e.g., 10\% - 90\%), denoted by $(\underline{\text{SOC}}, \overline{\text{SOC}})$:
\begin{align}\label{eq:SOC con}
\underline{\text{SOC}} \le \text{SOC}_{v}(t) \le \overline{\text{SOC}}. 
\end{align}

\subsubsection{Discharging power}
The discharging power, $p^{\text{EV-}}_{vt}$, depends on several quantities, including driving demand, driving style, regenerative breaking, auxiliaries' consumption (e.g., \cite{de2013plug}) and battery self-discharge. In this paper, $p^{\text{EV-}}_{vt}$ is estimated from historical time series of vehicles' SOC, as detailed in the case study section, hence providing a lumped description of all the underlying quantities impacting on it. It is worth highlighting that since the discharging power is assumed estimated directly from the vehicles' SOCs, it is not weighted by the efficiency in \eqref{eq:SOC model}.




\subsection{Identifying needs for charging infrastructure}
This key section of the paper explains how the number of chargers and their location are identified. Then, based on the number of chargers, the capital investment of the charging infrastructure is modeled.

\subsubsection{Single-port chargers}
It is first considered the case of single-port chargers. This case features an equal number of plugs and chargers (Fig.~\ref{fig:chargers}). 

The need for fast chargers (for slow chargers, the principles are identical and not repeated) in use at a specific grid node can be evaluated by coupling the information $f^\text{plugged}_{vt}$, telling whether a vehicle is connected to a fast charger, and $p_{nvt}$, telling its parking location. More specifically, the number of fast chargers in use at time interval $t$ at node $n$ is the sum over all vehicles of the product between $p_{nvt}$ and $f^\text{plugged}_{vt}$; by taking the maximum over time of this expression, one can determine the number of fast chargers required to meet the demand of fast chargers. Let $F^\text{chargers}_n$ and $S^\text{chargers}_n$ be the required number of fast and slow chargers, respectively, at node $n$. Based on the explanation above, they formally are:
\begin{subequations}
\begin{align}
F^\text{chargers}_n = \underset{t}{\text{max}} \left\{\sum_{v=1}^V p_{nvt}\cdot f^\text{plugged}_{vt}\right\}, && n=1, \dots, N \label{eq:number_of_chargers}\\
S^\text{chargers}_n = \underset{t}{\text{max}} \left\{\sum_{v=1}^V p_{nvt}\cdot s^\text{plugged}_{vt}\right\}, && n=1, \dots, N
\end{align}
For SPCs, the number of plugs for fast and slow chargers, denoted by $F^\text{plugs}_n$, $S^\text{plugs}_n$ respectively, are:
\begin{align}
    F^\text{plugs}_n = F^\text{chargers}_n\\
    S^\text{plugs}_n = S^\text{chargers}_n.
\end{align}\label{eq:SPCs}\end{subequations}

\subsubsection{Multi-port chargers}
With MPCs, a single charger can have have multiple plugs. Hence, the numbers of plugs and chargers now follow from different models. In particular, the number of plugs is evaluated considering the variables $f^\text{plugged}_{vt}, s^\text{plugged}_{vt}$, whereas the number of chargers is evaluated considering $f^\text{charge}_{vt}, s^\text{charge}_{vt}$, which tell how much power rating is used for charging the EVs. Using similar considerations as developed for the SPCs case, the number of plugs and number of chargers are:
\begin{subequations}
\begin{align}
F^\text{chargers}_n = \underset{t}{\text{max}} \left\{\sum_{v=1}^V p_{nvt}\cdot f^\text{charge}_{vt}\right\}, && n=1, \dots, N \\
S^\text{chargers}_n = \underset{t}{\text{max}} \left\{\sum_{v=1}^V p_{nvt}\cdot s^\text{plugged}_{vt}\right\}, && n=1, \dots, N. \\
F^\text{plugs}_n = \underset{t}{\text{max}} \left\{\sum_{v=1}^V p_{nvt}\cdot f^\text{plugged}_{vt}\right\}, && n=1, \dots, N \label{eq:number_of_fastPlugs} \\
S^\text{plugs}_n = \underset{t}{\text{max}} \left\{\sum_{v=1}^V p_{nvt}\cdot s^\text{plugged}_{vt}\right\}, && n=1, \dots, N
\end{align}\label{eq:MPCs}\end{subequations}

It is worth to highlight that if the solution of the MPCs problem is such that $f^\text{charge}_{vt} = f^\text{plugged}_{vt}$ and $s^\text{charge}_{vt} = s^\text{plugged}_{vt}$ for all $v$ and $t$, then the MPCs case reduces to the SPCs case. In this sense, the MPCs problem is a generalization of the SPCs case because it can lead to the same solution.

\subsubsection{Investment costs for the charging infrastructures}\label{sec:costs}
Based on the required numbers of plugs and chargers, we can estimate the capital cost of the charging infrastructure. The total investment cost is denoted by $J(\cdot)$, where notation $(\cdot)$ refers to the dependency of $J$ on the problem decision variables $f^\text{plugged}_{vt}$, $s^\text{plugged}_{vt}$, $f^\text{charge}_{vt}$, and $s^\text{charge}_{vt}$, not explicitly reported for compactness. It is:
\begin{subequations}
\begin{align}
J(\cdot) = J^F_{plugs} + J^F_{chargers} + J^S_{plugs} +J^S_{chargers} \label{eq:cost}
\end{align} 
where $J^F_{plugs}$, $J^F_{chargers}$ are the cost of fast-charging plugs and stations, and $J^S_{plugs}$ and $J^S_{chargers}$ are the cost of slow-charging plugs and stations. The components of \eqref{eq:cost} are as follows:
\begin{align}
J^F_{plugs} &= \sum_{n=1}^N F^\text{plugs}_n \cdot \text{cost}^F_{plugs}\\ 
J^S_{plugs} &= \sum_{n=1}^N S^\text{plugs}_n \cdot \text{cost}^S_{plugs} \\
J^F_{chargers} &= \sum_{n=1}^N F^\text{chargers}_n \cdot \text{cost}^F_{chargers}  \\
J^S_{chargers} &= \sum_{n=1}^N S^\text{chargers}_n \cdot \text{cost}^S_{chargers}
\end{align} \end{subequations}
where $\text{cost}^{F}_{plugs}, \text{cost}^{F}_{chargers}, \text{cost}^{S}_{plugs}, \text{cost}^{S}_{chargers}$ are the unitary cost of plugs and chargers for fast and slow charging.

\subsection{Nodal injections due to EVs charging demand and grid model}\label{sec:grid}
The problem formulation so far has focused on modeling the connection of EVs to chargers, their charging process, and how these reflect on the cost of the charging infrastructure. In this section, the charging demand of the EVs is used in a grid's load flow model to assess whether grid constraints are respected. These additional constraints are implemented in the planning problem with the specific objective of locating the chargers in the distribution grid without violation its operational limits. As load flow models are nonconvex, we resort to a linearized load flow based on sensitivity coefficients, as proposed in \cite{6473866,9243122,ISGT2021V2G,CIRED2021}, to retain the problem's tractability.

We denote the total charging demand for the EVs connected to node $n$ by $P^{\text{EV}}_{nt}$, and the associated reactive power demand by $Q^{\text{EV}}_{nt}$. These quantities are computed by coupling the information on the charging power of the individual EVs, $p^\text{EV+}_{vt}$ in \eqref{eq:ev_charging_power}, with their parking location, $p_{nvt}$. Formally, they are:
\begin{subequations}\label{eq:ev_nodal_injections}
\begin{align}
& P^{\text{EV}}_{nt} = \sum_{v=1}^{V} p_{nvt}\cdot  p^{\text{EV+}}_{vt} && \forall t \text{ and } n\\
& Q^{\text{EV}}_{nt} = \sum_{v=1}^{V} p_{nvt}\cdot  q^{\text{EV+}}_{vt} && \forall t \text{ and } n.
\end{align}
\end{subequations}

The nodal power injections at the various nodes of the distribution grid are given by the total charging demand of the EVs in \eqref{eq:ev_nodal_injections} along with conventional demand and local distributed generation. Conventional demand and distributed generation is modeled in terms of net demand, denoted by $P^{\text{net}}_{nt}$, given by the difference between the two. The net demand is an input of the problem. Nodal active and reactive power injections read as:
\begin{subequations}\label{eq:nodal_injections}
\begin{align}
& P^{\text{node}}_{nt} = \sum_{v \in \mathcal{V}} p_{nvt}\cdot   p^{\text{EV+}}_{vt} + P^{{\text{net}}}_{nt} \label{eq:Pnode} \\
& Q^{\text{node}}_{nt} = \sum_{v \in \mathcal{V}} p_{nvt}\cdot  q^{\text{EV+}}_{vt} + Q^{\text{net}}_{nt} \label{eq:Qnode}.
\end{align}
\end{subequations}
Nodal power injections are assumed voltage independent.

Nodal voltage magnitudes $v_{tn}$, line current magnitudes $i_{tl}$ in lines $l=1, \dots,L$, and apparent power flow at the substation transformer $S_{t0}$ are denoted with the following notation
\begin{subequations}\label{eq:gridmodel}
\begin{align}
& v_{tn} = f_n\left(P^\text{node}_{t1},\dots, P^\text{node}_{tN}, Q^\text{node}_{t1},\dots, Q^\text{node}_{tN}\right) \label{volt}\\
& i_{tl} = g_l\left(P^\text{node}_{t1},\dots, P^\text{node}_{tN}, Q^\text{node}_{t1},\dots, Q^\text{node}_{tN}\right) \label{curr}\\
& S_{t0} = h_l\left(P^\text{node}_{t1},\dots, P^\text{node}_{tN}, Q^\text{node}_{t1},\dots, Q^\text{node}_{tN}\right)\label{power}
\end{align}
which highlights the dependency between grid quantities and nodal injections through the functions $f_n$, $g_l$, and $h_l$. The problem dependency on the admittance matrix (i.e., topology and cables' characteristics), and slack bus voltage are omitted from this notation for simplicity. Using the notion of sensitivity coefficients, functions $f_n$, $g_l$, and $h_l$ can be expressed as a linear function of the nodal power injections and linearization points.

Operational constraints of the distribution grid refer to voltage magnitude within prescribed limits $(\underline{v}, \overline{v})$, currents in the lines below the lines' ampacities $\overline{i}_l$, and power flow at the substation transformer less than its rating $S_{t0}$. These reads as:
\begin{align}
& \underline{v} \le v_{tn} \le \overline{v} && \forall t \text{ and } n\\
& |i_{tl}| \le \overline{i}_l  && \forall t \text{ and } l\\
& S_{t0} \le \overline{S}_0 && \forall t. \label{eq:nodalPowerConst1}
\end{align}
In addition to these constraints, we require nodal injections to be below the apparent power limit of the node, $S_n$:
\begin{align}
    {(P^{\text{node}}_{nt})}^2 + {(Q^{\text{node}}_{nt})}^2 \le (S_n)^2. \label{eq:LV_transformer1}
\end{align}\end{subequations}
Constraint \eqref{eq:LV_transformer1} is useful in the case of apparatus with apparent power limitations connected at the nodes, such as nodes hosting substation step-down transformers.

\subsection{Planning problem}
The planning problem consists in finding binary variables
\begin{align}
\boldsymbol{x} &= [f^{\text{charge}}_{11}, \dots, f^{\text{charge}}_{VT}, s^{\text{charge}}_{11}, \dots, s^{\text{charge}}_{VT}] \\
\boldsymbol{y} &= [f^{\text{plugged}}_{11}, \dots, f^{\text{plugged}}_{VT}, s^{\text{plugged}}_{11}, \dots, s^{\text{plugged}}_{VT}]
\end{align}
that minimize the capital investment for the EV charging infrastructure while subject to grid constraints.

To avoid that the problem solution depend on the initial SOC values in \eqref{eq:SOC model}, we choose to set them as problem variables, denoted by:
\begin{align}
\boldsymbol{z} = [\text{SOC}_{1}(0), \dots, \text{SOC}_{V}(0)] \in \mathbb{R}^V.
\end{align}
Besides, the final SOC should be larger than or equal to the initial one to avoid benefiting from the initial energy stock:
\begin{align}
    \text{SOC}_{v}(T) \ge  \text{SOC}_{v}(0), && \text{for all } v.\label{eq:SOC con1}
\end{align}
In this way, the planning problem accounts for the charging demand of the vehicles, regardless of their specific initial conditions. 

The planning problem is formulated as a constrained economic optimization. Its formulation reads as:
\begin{subequations}
\begin{align}
    \underset{\boldsymbol{x,y} \in \{0,1\}^{V \times T}, \boldsymbol{z} \in \mathbb{R}^V }{\text{min}} \left\{ J(\cdot) \right\}
\end{align}
subject to the following constraints:
\begin{align}
& \text{Plugged-in only if parked \eqref{eq:pluggedinonlyifparked}} && \text{$\forall t$ and $v$} \\
& \text{Charge only if plugged-in \eqref{eq:chargeonlyifparked}} && \text{$\forall t$ and $v$} \\
& \text{EV charging power \eqref{eq:ev_charging_power}} && \text{$\forall t$ and $v$}\\
& \text{SOC model and constraints \eqref{eq:SOC model}, \eqref{eq:SOC con}, \eqref{eq:SOC con1}} && \text{$\forall t$ and $v$}\\
& \text{Nodal injections \eqref{eq:ev_nodal_injections} and \eqref{eq:nodal_injections}} && \text{$\forall t$ and $n$} \\
& \text{Linear grid models and constraints \eqref{eq:gridmodel}} \\
& \text{Chargers and plugs number model:} \nonumber \\
& \text{~~~~ \eqref{eq:SPCs} for SPCs, or \eqref{eq:MPCs} for MPCs}
\end{align}\label{eq:optproblem}\end{subequations}

\subsection{Problem properties and approximations}
Problem \eqref{eq:optproblem} is nonlinear due to the set maximum in \eqref{eq:SPCs}-\eqref{eq:MPCs}, the point-wise maximum in \eqref{eq:disconnConsSlow} (a new constraint, explained in the next section), and the quadratic expression in \eqref{eq:LV_transformer1}. Suitable reformulations or approximations of these constraints are now discussed to render the problem linear. The set maximum, here denoted by $\bar{v} = \text{max} \{v_t, t=1,\dots, T\}$ for convenience, is replaced by $T$ linear inequalities $\bar{v} \ge v_t$ for all $t$. As the problem \eqref{eq:optproblem} entails \emph{minimizing} expressions of the same kind as $\bar{v}$, this reformulation holds as exact. The point-wise maximum, $a^+ = \text{max}(a, 0)$, is replaced by 2 inequalities, $a^+ \ge a$ and $a^+ \ge 0$ and can be used to replace convex constraints in the form of $\text{max}(a, 0) \le \bar{a}$ with linear ones.

Finally, the apparent power constraint in \eqref{eq:LV_transformer1}, now denoted by $ P^2 + Q^2 \le S^2$ for simplicity, is approximated by replacing the reactive power with an upper bound $\overline{Q} = S \cdot {\overline{\sin} \phi}$; since $Q \le \overline{Q}$, it follows that:
\begin{subequations}
\begin{align}
    & P^2 + Q^2 \le P^2 + \overline{Q}^2 \le S^2\\
    & P^2 \le S^2 - \overline{Q}^2 = S^2 - S^2 \cdot \overline{\sin}^2 \phi = S^2 \underline{\cos}^2 \phi\\
    & P \ge -S \cdot \underline{\cos} \phi \text{~~and~~} P \le S \cdot \underline{\cos} \phi.\label{eq:linearapparentpower}
\end{align}\end{subequations}
In summary, the original quadratic constraint is replaced by two linear inequalities, \eqref{eq:linearapparentpower}, with $\underline{\cos} \phi$ as a estimated lower bound of the load power factor. An alternative approach is to approximate the convex set  \eqref{eq:LV_transformer1} with linear inequalities (e.g., \cite{nick2014optimal}), typically preferrable when reactive power is an explicit control variable of the problem. With these equivalent formulations and approximation, it is possible to write the optimization problem as a mixed integer linear problem (MILP).

\section{Modeling drivers connection and disconnection preferences}

It has been said that $f^\text{plugged}_{vt}$ and $s^\text{plugged}_{vt}$ (denoting if an EV is plugged into a charger) can be active only when an EV is parked. However, there is more. Because plugging an EV into a charging column is an operation performed by the drivers, their availability to plug and unplug an EV should also be modelled. For example, a person driving home in the evening and using a public charging column might prefer to plug their EV when arriving rather than waiting for a busy charger to be available and come back in the middle of the night to plug it. In this context, by adding additional constraints on $f^\text{plugged}_{vt}$ and $s^\text{plugged}_{vt}$, we model two scenarios to capture different levels of drivers' flexibility for plugging and unplugging their EVs. To explain these constraints, we refer to the case study analyzed in this paper (described in detail in the next session) that considers a home-work commute, where EVs are used in the morning, parked in the central part of the day, used again in the afternoon, and finally parked overnight (\footnote{We recall that this is input information and is encoded in parameters $p_{nvt}$.}). The constraints to model drivers' flexibility are discussed in the rest of this section.

\subsection{Modeling connection to and disconnection from chargers}
\begin{figure}[h!]
    \centering
    \includegraphics[width=0.8\columnwidth]{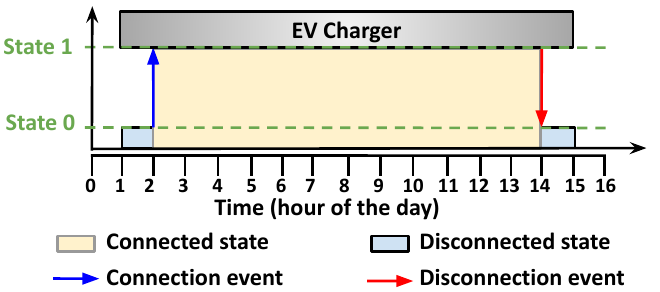}
    \caption{Example of the connection-state variable ($f^\text{plugged}_{vt}$ or $s^\text{plugged}_{vt}$) and connection and disconnection events (blue and red arrows), corresponding to the raising and falling edges of the connection state, respectively.}
    \label{fig:ConnDisconn}
\end{figure}

Before describing the drivers' flexibility scenarios, the models to compute the connections to and disconnections from chargers are explained. For fast chargers, let binary variables $c^{f}_{vt}, d^{f}_{vt}$ denote the events when EV $v$ is connected to and disconnected from a charger, respectively, and similarly for slow chargers, with variables $c^{s}_{vt}$ and $d^{s}_{vt}$. In these variables, the logical state 1 denotes a connection or a disconnection, and 0 no event. Connections and disconnections are modeled by detecting rising and falling edges of $f^\text{plugged}_{vt}$ and $s^\text{plugged}_{vt}$ (Fig. \ref{fig:ConnDisconn}). Formally, this is as (with max as the point-wise maximum):
\begin{subequations}
\begin{align}
c^{f}_{vt} = {\text{max}} \left(f^\text{plugged}_{vt} - f^\text{plugged}_{v(t-1)}, 0\right) && \text{$\forall t$ and $v$} \label{eq:connConsFast}\\
d^{f}_{vt} = {\text{max}} \left(f^\text{plugged}_{v(t-1)} - f^\text{plugged}_{vt}, 0\right) && \text{$\forall t$ and $v$} \label{eq:disconnConsFast}\\
c^{s}_{vt} = {\text{max}} \left(s^\text{plugged}_{vt} - s^\text{plugged}_{v(t-1)}, 0\right) && \text{$\forall t$ and $v$} \label{eq:connConsSlow}\\
d^{s}_{vt} = {\text{max}} \left(s^\text{plugged}_{v(t-1)} - s^\text{plugged}_{vt}, 0\right) && \text{$\forall t$ and $v$}. \label{eq:disconnConsSlow}\end{align}\end{subequations}

\subsection{Drivers flexibility scenarios}\label{scenarios}
Let the time interval $(\tau^{(1)}_v, \tau^{(2)}_v)$ denote the overnight parking stay, and $(\tau^{(3)}_v, \tau^{(4)}_v)$ the parking stay during the central hours of the day for vehicle $v$. The two drivers' flexibility scenarios are as follows.

\textbf{Scenario A (stiff drivers)}: \emph{In both parking intervals, drivers plug their EVs to a charger only at the arrival time, and unplug them only at the departure time.} In other words, drivers let their vehicles plugged into a charger whenever their EVs is parked; for their comfort, plugging and unplugging happen only at the arrival and departing times. Formally, this is implemented by enforcing no connection outside the initial parking time interval (for both fast and slow chargers)
\begin{subequations}
\begin{align}
& c^{f}_{vt} \le 0  && \text{for all $t$ except $t=\tau^{(1)}_v$ and $t=\tau^{(3)}_v$} \\ 
& c^{s}_{vt} \le 0  && \text{for all $t$ except $t=\tau^{(1)}_v$ and $t=\tau^{(3)}_v$},
\end{align}
and no disconnection outside the final parking time interval
\begin{align}
& d^{f}_{vt} \le 0 && \text{for all $t$ except $t=\tau^{(2)}_v$ and $t=\tau^{(4)}_v$} \\ 
& d^{s}_{vt} \le 0 && \text{for all $t$ except $t=\tau^{(2)}_v$ and $t=\tau^{(4)}_v$}.
\end{align}\label{eq:scenarioa}\end{subequations}

\textbf{Scenario B (flexible drivers)}: \emph{For overnight parking, drivers plug their EVs to a charger only at the arrival time, and unplug them only at the departure time; for central parking hours, drivers allow one disconnection}.  In other words, drivers allow one disconnection to give to others the possibility of using that charging spot. This is implemented by enforcing no connection outside the initial parking time for the overnight time interval
\begin{subequations}
\begin{align}
& c^{f}_{vt} \le 0  && \text{for all $t$ except $t=\tau^{(1)}_v$} \\ 
& c^{s}_{vt} \le 0  && \text{for all $t$ except $t=\tau^{(1)}_v$},
\end{align}
and up to one disconnection in the central parking hours
\begin{align}
\sum_{t=\tau_3}^{\tau_4} d^{f}_{vt} \le 1, \\
\sum_{t=\tau_3}^{\tau_4} d^{s}_{vt} \le 1.
\end{align}\label{eq:scenariob}\end{subequations}

\subsection{Implementing the scenarios}
Scenarios are implemented by adding either \eqref{eq:scenarioa} or \eqref{eq:scenariob} to optimization problem \eqref{eq:optproblem}. A comparative analysis of these 2 scenarios is performed in the results section to evaluate the impact of drivers' flexibility on the problem solution.

\section{Case study}\label{sec:case_study}
This section describes the case study to which the proposed planning method is applied with the objective of exemplifying how the pieces of input information are computed.

The adopted study is reasonably guessed to reproduce a real possible scenario. It is worth remarking that the contributions of this paper do not depend on this specific case study; in particular, input information can always be tuned or changed as a function of the specific situation to model, on the basis of, for example, information from the distribution grid operator and urban planner.

\subsection{Number of EVs and driving demand}
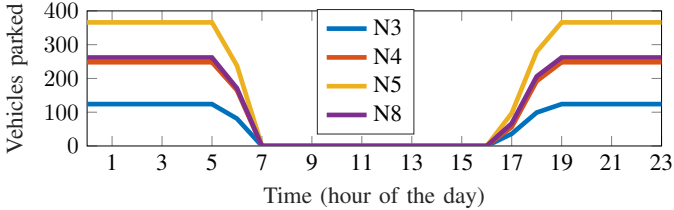
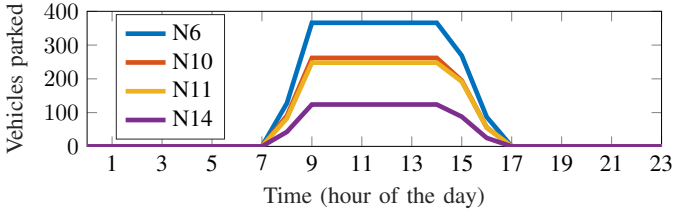
\begin{figure}[h!]
     \begin{subfigure}[b]{0.5\textwidth}
          \centering
          \resizebox{\linewidth}{!}{
%
%
\definecolor{mycolor1}{rgb}{0.00000,0.44700,0.74100}%
\definecolor{mycolor2}{rgb}{0.85000,0.32500,0.09800}%
\definecolor{mycolor3}{rgb}{0.92900,0.69400,0.12500}%
\definecolor{mycolor4}{rgb}{0.49400,0.18400,0.55600}%
\begin{tikzpicture}

\begin{axis}[%
width=3.4in,
height=0.8in,
at={(0.758in,0.481in)},
scale only axis,
xmin=1,
xmax=24,
xtick={2,4,6,8,10,12,14,16,18,20,22,24},
xticklabels={{1},{3},{5},{7},{9},{11},{13},{15},{17},{19},{21},{23}},
xlabel style={font=\color{white!15!black}},
xlabel={Time (hour of the day)},
ymin=0,
ymax=400,
ylabel style={font=\color{white!15!black}},
ylabel={Vehicles parked},
axis background/.style={fill=white},
legend style={at={(0.4,0.085)}, anchor=south west, legend cell align=left, align=left, draw=white!15!black}
]
\addplot [color=mycolor1, line width=2.0pt]
  table[row sep=crcr]{%
1	124\\
2	124\\
3	124\\
4	124\\
5	124\\
6	124\\
7	81\\
8	0\\
9	0\\
10	0\\
11	0\\
12	0\\
13	0\\
14	0\\
15	0\\
16	0\\
17	0\\
18	36\\
19	99\\
20	124\\
21	124\\
22	124\\
23	124\\
24	124\\
};
\addlegendentry{N3}

\addplot [color=mycolor2, line width=2.0pt]
  table[row sep=crcr]{%
1	248\\
2	248\\
3	248\\
4	248\\
5	248\\
6	248\\
7	165\\
8	0\\
9	0\\
10	0\\
11	0\\
12	0\\
13	0\\
14	0\\
15	0\\
16	0\\
17	0\\
18	56\\
19	192\\
20	248\\
21	248\\
22	248\\
23	248\\
24	248\\
};
\addlegendentry{N4}

\addplot [color=mycolor3, line width=2.0pt]
  table[row sep=crcr]{%
1	366\\
2	366\\
3	366\\
4	366\\
5	366\\
6	366\\
7	238\\
8	0\\
9	0\\
10	0\\
11	0\\
12	0\\
13	0\\
14	0\\
15	0\\
16	0\\
17	0\\
18	97\\
19	279\\
20	366\\
21	366\\
22	366\\
23	366\\
24	366\\
};
\addlegendentry{N5}

\addplot [color=mycolor4, line width=2.0pt]
  table[row sep=crcr]{%
1	262\\
2	262\\
3	262\\
4	262\\
5	262\\
6	262\\
7	171\\
8	0\\
9	0\\
10	0\\
11	0\\
12	0\\
13	0\\
14	0\\
15	0\\
16	0\\
17	0\\
18	67\\
19	206\\
20	262\\
21	262\\
22	262\\
23	262\\
24	262\\
};
\addlegendentry{N8}

\end{axis}
\end{tikzpicture}%
}
          \caption{Cluster 1}
          \label{nightHour_parking}
     \end{subfigure}
     \begin{subfigure}[b]{0.5\textwidth}
          \centering
          \resizebox{\linewidth}{!}{
%
%
\definecolor{mycolor1}{rgb}{0.00000,0.44700,0.74100}%
\definecolor{mycolor2}{rgb}{0.85000,0.32500,0.09800}%
\definecolor{mycolor3}{rgb}{0.92900,0.69400,0.12500}%
\definecolor{mycolor4}{rgb}{0.49400,0.18400,0.55600}%
\begin{tikzpicture}

\begin{axis}[%
width=3.4in,
height=0.8in,
at={(0.758in,0.481in)},
scale only axis,
xmin=1,
xmax=24,
xtick={2,4,6,8,10,12,14,16,18,20,22,24},
xticklabels={{1},{3},{5},{7},{9},{11},{13},{15},{17},{19},{21},{23}},
xlabel style={font=\color{white!15!black}},
xlabel={Time (hour of the day)},
ymin=0,
ymax=400,
ylabel style={font=\color{white!15!black}},
ylabel={Vehicles parked},
axis background/.style={fill=white},
legend style={at={(0.05,0.979)}, anchor=north west, legend cell align=left, align=left, draw=white!15!black}
]
\addplot [color=mycolor1, line width=2.0pt]
  table[row sep=crcr]{%
1	0\\
2	0\\
3	0\\
4	0\\
5	0\\
6	0\\
7	0\\
8	0\\
9	128\\
10	366\\
11	366\\
12	366\\
13	366\\
14	366\\
15	366\\
16	269\\
17	87\\
18	0\\
19	0\\
20	0\\
21	0\\
22	0\\
23	0\\
24	0\\
};
\addlegendentry{N6}

\addplot [color=mycolor2, line width=2.0pt]
  table[row sep=crcr]{%
1	0\\
2	0\\
3	0\\
4	0\\
5	0\\
6	0\\
7	0\\
8	0\\
9	91\\
10	262\\
11	262\\
12	262\\
13	262\\
14	262\\
15	262\\
16	195\\
17	56\\
18	0\\
19	0\\
20	0\\
21	0\\
22	0\\
23	0\\
24	0\\
};
\addlegendentry{N10}

\addplot [color=mycolor3, line width=2.0pt]
  table[row sep=crcr]{%
1	0\\
2	0\\
3	0\\
4	0\\
5	0\\
6	0\\
7	0\\
8	0\\
9	83\\
10	248\\
11	248\\
12	248\\
13	248\\
14	248\\
15	248\\
16	192\\
17	56\\
18	0\\
19	0\\
20	0\\
21	0\\
22	0\\
23	0\\
24	0\\
};
\addlegendentry{N11}

\addplot [color=mycolor4, line width=2.0pt]
  table[row sep=crcr]{%
1	0\\
2	0\\
3	0\\
4	0\\
5	0\\
6	0\\
7	0\\
8	0\\
9	43\\
10	124\\
11	124\\
12	124\\
13	124\\
14	124\\
15	124\\
16	88\\
17	25\\
18	0\\
19	0\\
20	0\\
21	0\\
22	0\\
23	0\\
24	0\\
};
\addlegendentry{N14}

\end{axis}
\end{tikzpicture}%
} 
          \caption{Cluster 2}
          \label{dayHour_parking}
     \end{subfigure}
    \caption{Number of EVs' parked at different grid nodes under two clusters during the day and night hours.}\label{s_vt}
\end{figure}

It is considered a home-work commute where EVs are used in the morning, parked in the central part of the day, used again in the afternoon, and parked overnight at the origin node. The different parking (and charging) locations correspond to different nodes of the grid. The residential nodes where EVs are parked overnight are indicated as the green nodes (\say{Cluster 1}) in Fig.~1, whereas the destination nodes are the purple nodes (\say{Cluster 2}). In total, there are 1'000 EVs in this grid. This value is chosen based on the rating of this power grid, and it is in line with other studies (e.g., \cite{en8031760, CIRED2021}). The origin and destination nodes of the EVs are assigned randomly and uniformly to all nodes hosting EVs. The number of EVs parked during the night and central hours are shown in Fig.~\ref{nightHour_parking} and ~\ref{dayHour_parking}, respectively. The EVs' morning departures and arrivals are sampled from uniform distributions with values between hours 5-8 and 8-11, respectively; evening departures and arrivals are sampled from uniform distributions with values between hours 14-18 and 17-21. Based on the information presented so far in this section, variables $p_{nvt}$ are built. It is worth highlighting that this is an input of the problem, and other methods (including using real data) can be used. The total energy demand for driving of an EV is estimated by comparing the daily starting and final SOC of EVs from \cite{testanevrepo}. The discharging power, $p^{\text{EV-}}_{vt}$, necessary to model the SOC evolution in \eqref{eq:SOC model}, is assumed piecewise constant, strictly positive during the intervals when the vehicle drives, zero when the EV is parked, and such that the associated energy demand amounts to the quantity estimated above.  We have considered a nominal energy capacity of the EVs' batteries of 16 kWh for all EVs.

The proposed analyses are for an optimization horizon of 24 hours, which is assumed to capture a typical day of driving demand. The resolution of the input time series is 1 hour.

\subsection{Chargers ratings and prices}
We consider fast and slow chargers with kVA ratings of 50 kVA and 2.4 kVA, respectively, and power factors of 0.9. Their costs is assumed to be 20'000€ and 800€, in line with the existing technical literature \cite{PriceEUreport,RMIreport, slowChargerPrice} (although some price volatility might exist due to different regions, operators, and need for labor). The price of the charging plugs is assumed to be 15\% of the price of a single-port charger.

\subsection{Distribution grid and demand}
\begin{figure}[!h]
    \centering
    \includegraphics[width=0.85\columnwidth]{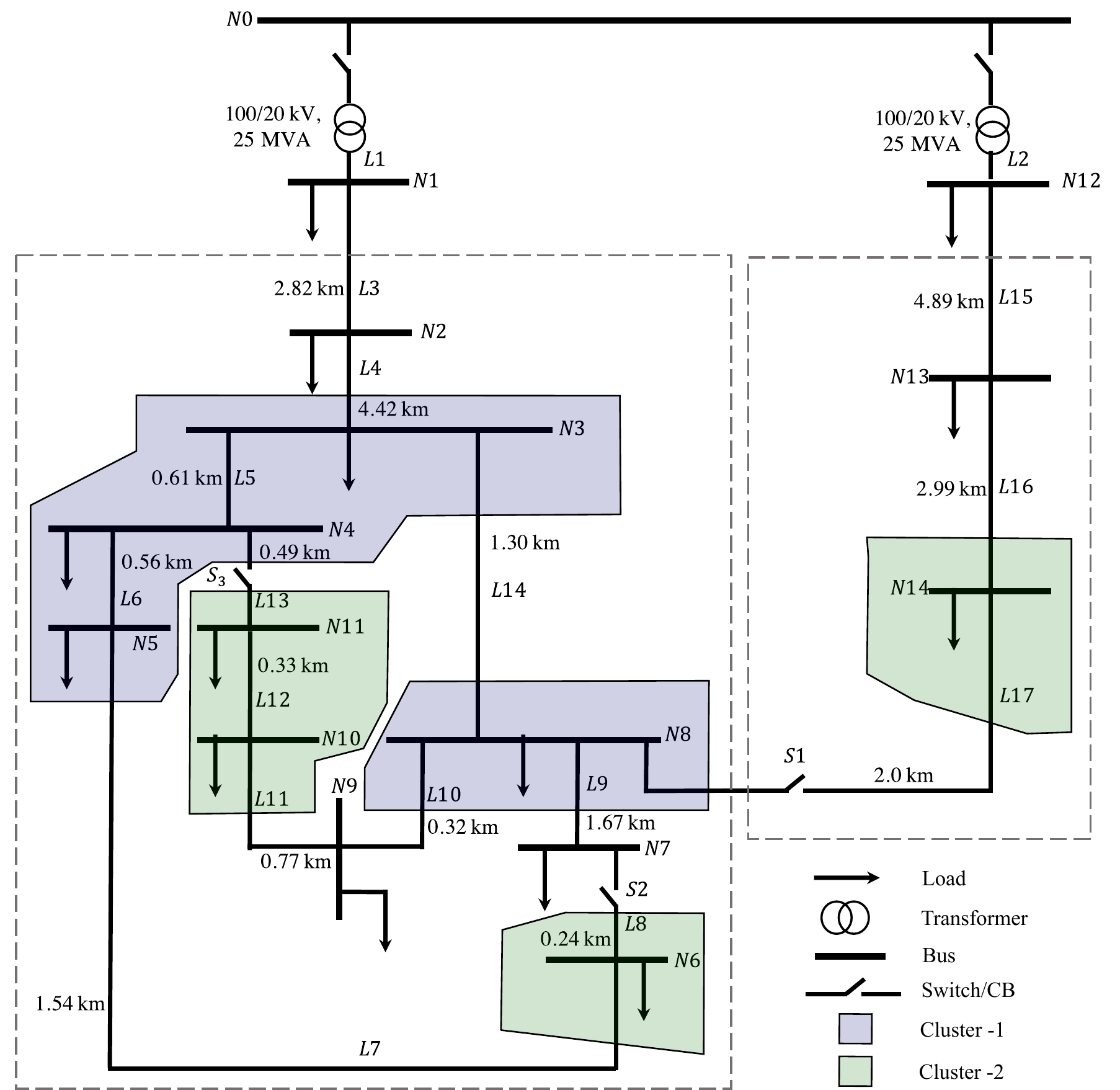}
    \caption{Topology of the CIGRE European MV distribution network benchmark for residential system \cite{CIGREREF}.}
    \label{fig:CIGREMV}
\end{figure}

It is considered the European version of the 14-bus CIGRE benchmark grid for medium voltage (MV) systems \cite{CIGREREF} (Fig.~\ref{fig:CIGREMV}). The low-voltage (LV) grids connected at the MV grid nodes are modeled in terms of their aggregated power. This modeling accounts for constraints of the rated power of the MV/LV substation transformer through \eqref{eq:LV_transformer1}, and assumes that there are no violations of voltage levels and line ampacities in the LV grid.
The MV grid is modeled as a single-phase equivalent assuming transposed conductors and balance loads. The demand of the grid is simulated considering the load profile proposed in \cite{CIGREREF}, scaled according to the rated power of each node (Table~\ref{tab:tab1}). At this stage, no distributed renewable generation is considered. The reactive power of the nodal injections is modeled assuming a constant power factor (Table~\ref{tab:tab1}). Statutory voltage levels are 1 $\pm$ 3$\%$ per unit of the base voltage (20~kV). Line ampacities are according to the conductor diameter. The sensitivity coefficients for the linearized grid model are computed once for the nominal demand profiles; one could compute successive linearizations to improve the linear estimates.

\begin{table}[!h]
\caption{Nodal nominal demand and power factors}\label{tab:tab1}
\begin{center}
{\small
\begin{tabular}{|c||c|c|c|}\hline
Node & Apparent Power [kVA]  & Power factor & Cluster\\
\hline
1 & 15'300 & 0.98 & - \\
3 & 285 & 0.97 & 1 \\ 
4 & 445 & 0.97 & 1 \\
5 & 750 & 0.97 & 1  \\
6 & 565 & 0.97 & 2\\
8 & 605 & 0.97 & 1  \\
10 & 490 & 0.97 & 2  \\ 
11 & 340 & 0.97 & 2\\ 
12 & 15'300 & 0.98 & -\\
14 & 215 & 0.97 & 2\\
\hline
\end{tabular}}
\end{center}
\end{table}

\section{Results and Discussions}
The proposed planning method is applied to the case study illustrated in the former section. Results are now discussed. The discussion is organized as follows. First, a subset of the problem variables is shown to highlight specific properties of the solution. Then, EVs charging patterns and active constraints of the optimization problem are discussed. A comparative analysis that includes SPCs, MPCs, and flexible driver scenarios is then presented. Finally, a sensitivity analysis of the results with respect to increasing levels of driving demand is performed to verify how these impact the planning solution.


\subsection{Exemplifying decision variables $s^{\text{plugged}}_{vt}$ and $s^{\text{charge}}_{vt}$}
\begin{figure}[!h]
     \begin{subfigure}[b]{0.5\textwidth}
          \centering
          \resizebox{\linewidth}{!}{\input{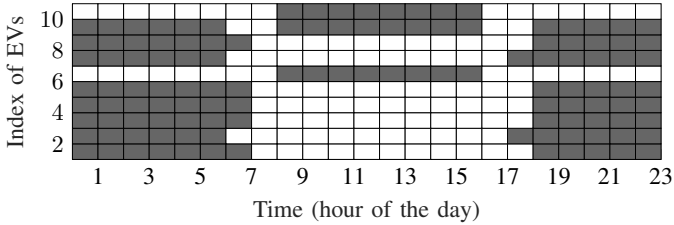}}
          \caption{}
          \label{s_plugged}
     \end{subfigure}\vspace*{3pt}
     \begin{subfigure}[b]{0.5\textwidth}
          \centering
          \resizebox{\linewidth}{!}{\input{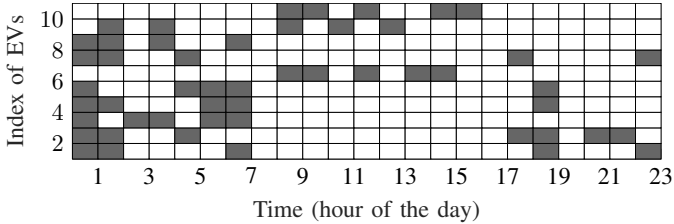}}
          \caption{}
          \label{s_charge}
     \end{subfigure}
    \caption{Variables $s^{\text{plugged}}_{vt}$ (a) and $s^{\text{charge}}_{vt}$ (b), showing the connection and charging state, respectively, for 10 sample EVs. Grey filling is 1, white 0. }\label{s_charge_plug_variables}
\end{figure}

Fig.~\ref{s_charge_plug_variables} shows variables $s^{\text{plugged}}_{vt}$ and $s^{\text{charge}}_{vt}$ for ten sample EVs in Scenario A and MPCs in order to illustrate their meaning. It shows that, i), vehicles are mostly connected to the chargers. This is in line with the definition of Scenario A, which foresees EVs connected to the chargers whenever they are parked; ii), the planning algorithm arbitrages the  charging of plugged EVs. This is done to respect grid constraints, ensure the EVs have correct SOC levels throughout the day and attain a minimum investment cost, as dictated by the problem cost function. We can thus infer that arbitraging the charge is beneficial to reducing the number of chargers, as explained hereafter.

\subsection{Charging infrastructure requirements}
The number of required chargers and plugs for SPCs, MPCs, and the two scenarios for driver's flexibility scenarios A and B are reported in Table~\ref{tab:caseAll}. For SPCs, the number of plugs is not indicated as it is the same as the number of chargers. It is possible to derive the following findings.

\textbf{Finding 0}: fast chargers are not required. In the considered case study, cheaper distributed slow chargers were enough to satisfy the charging demand.

\textbf{Finding 1}: moving from stiff to flexible drivers (from Scenario A to B) for both SPCs and MPCs attains smaller numbers of chargers and plugs. This is explained by the fact that increasing the availability of the drivers to plug/unplug their EVs leads to better utilization of the charging infrastructure, ultimately requiring fewer chargers to satisfy the same charging demand. 

\textbf{Finding 2}: implementing MPCs requires less chargers and more plugs. As the MPCs problem is a generalization of the SPCs' and the problem aims at finding the economic minimum, we can infer that MPCs are conducive to lower infrastructure costs (as empirically confirmed by the economic analysis reported in the next paragraph). 

\textbf{Finding 3}: different cases (scenarios and MPCs/SPCs) might entail a different spatial distribution of the chargers. Chargers of Scenario A/SPCs are nearly equally split between Cluster 1 and 2's nodes, whereas Scenario B/SPCs place more chargers in Cluster 2 (overnight stay). This might be due to longer overnight parking stays, which increase the possibility of arbitrating the charge between a larger group of vehicles, leading to a more efficient use of the charging infrastructure.

\begin{table}[!h]
    \caption{Number of slow chargers and plugs}
    \centering
    \tabcolsep 4pt
    \begin{tabular} {|c||c|c|c|c|c|c|}\hline
        & \multicolumn{3}{|c|}{ Scenario A} & \multicolumn{3}{|c|}{Scenario B}\\ \cline{2-7}
       Node & \multicolumn{2}{|c|}{MPCs} &  \multicolumn{1}{|c|}{SPCs} &  \multicolumn{2}{|c|}{MPCs} &  \multicolumn{1}{|c|}{SPCs} \\\cline{2-7}
             & Chargers & Plugs & Chargers & Chargers & Plugs & Chargers \\\cline{1-7}
3 & 44 & 124 & 41 & 36 & 111 & 32 \\
4 & 70 & 197 & 130 & 50 & 173 & 109 \\
5 & 119 & 287 & 215 & 89 & 271 & 242 \\
6 & 60 & 115 & 155 & 87 & 110 & 124 \\
8 & 96 & 234 & 133 & 75 & 212 & 143 \\
10 & 29 & 60 & 130 & 53 & 63 & 119 \\
11 & 41 & 103 & 130 & 56 & 64 & 71 \\
14 & 3 & 10 & 87 & 14 & 14 & 46 \\
\hline
Total & 462 & 1130 & 1021 & 460 & 1018 & 886 \\
Cluster1 & 329 & 842 & 519 & 250 & 767 & 526 \\
Cluster2 & 133 & 288 & 502 & 210 & 251 & 360 \\
\hline
    \end{tabular}
    \label{tab:caseAll}
\end{table}

\subsection{Economics}
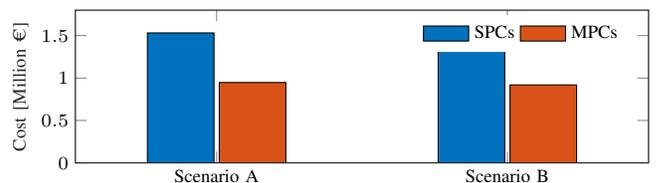
\begin{figure}[!h]
\centering
\scriptsize
%
%
\definecolor{mycolor1}{rgb}{0.00000,0.44700,0.74100}%
\definecolor{mycolor2}{rgb}{0.85000,0.32500,0.09800}%
\begin{tikzpicture}

\begin{axis}[%
width=3.0in,
height=0.80in,
at={(0.758in,0.481in)},
scale only axis,
bar shift auto,
xmin=0.514285714285714,
xmax=2.48571428571429,
xtick={1,2},
xticklabels={{Scenario A},{Scenario B}},
ymin=0,
ymax=1.800000,
ylabel style={font=\color{white!15!black}},
ylabel={Cost [Million €]},
axis background/.style={fill=white},
legend style={legend pos= north east, legend columns=3, legend cell align=left, align=left, draw=none}
]
\addplot[ybar, bar width=0.229, fill=mycolor1, draw=black, area legend] table[row sep=crcr] {%
1	1.531500\\
2	1.329000\\
};
\addplot[forget plot, color=white!15!black] table[row sep=crcr] {%
0.514285714285714	0\\
2.48571428571429	0\\
};
\addlegendentry{SPCs}

\addplot[ybar, bar width=0.229, fill=mycolor2, draw=black, area legend] table[row sep=crcr] {%
1	0.947250\\
2	0.919050\\
};
\addplot[forget plot, color=white!15!black] table[row sep=crcr] {%
0.514285714285714	0\\
2.48571428571429	0\\
};
\addlegendentry{MPCs}

\end{axis}

\end{tikzpicture}%
\caption{Cost of the four cases. }\label{costPlot}
\end{figure}

The cost achieved by the four analyzed cases are summarized in Fig.~\ref{costPlot}. Two additional findings are derived.

\textbf{Finding 4}: Implementing MPCs and flexible drivers (Scenario B) are beneficial from a cost perspective. 

\textbf{Finding 5}: The cost savings achieved by MPCs are significantly more substantial than flexible drivers. Choosing MPCs over SPCs achieves a cost reduction of 38\% and 30\% in Scenario A and B, respectively. Implementing flexible drivers (Scenario B) achieves a cost reduction of 13\% and 3\% for SPCs and MPCs, respectively. The important implication here is that a technological solution obtains a better effect than promoting consumer behavior change.

\subsection{Nodal injections}

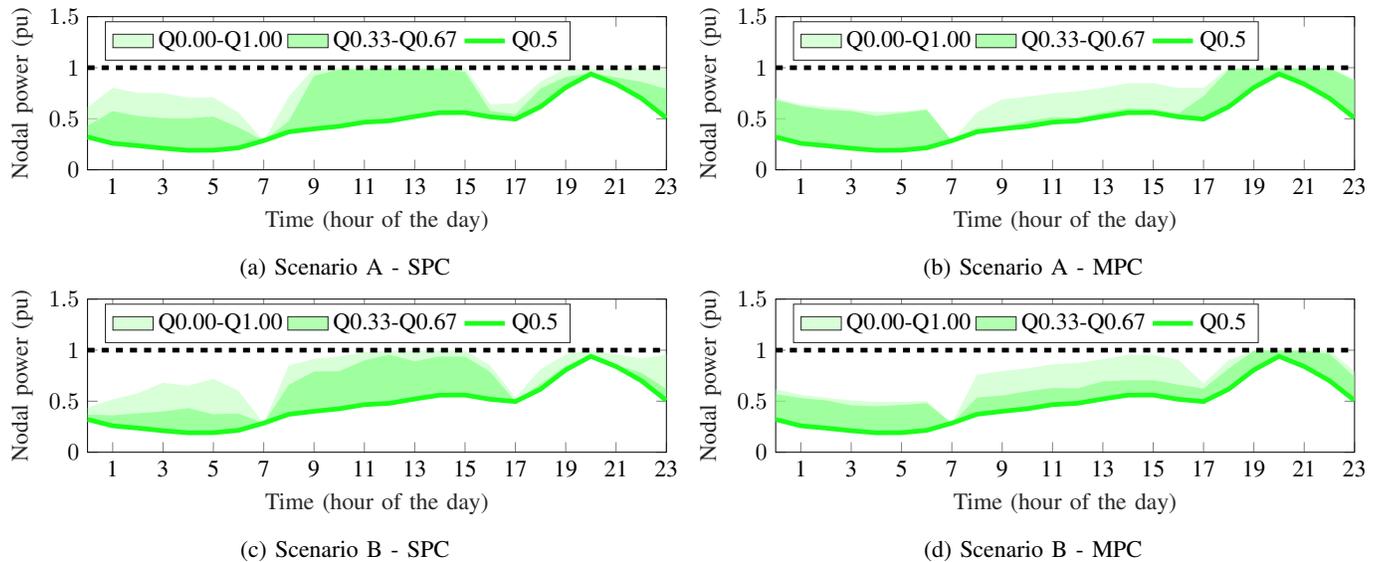
\begin{figure*}[!h]
     \begin{subfigure}[b]{0.5\textwidth}
          \centering
          \resizebox{\linewidth}{!}{
%
%
\begin{tikzpicture}

\begin{axis}[%
width=3.4in,
height=0.9in,
at={(0.758in,0.481in)},
scale only axis,
xmin=1,
xmax=24,
xtick={2,4,6,8,10,12,14,16,18,20,22,24},
xticklabels={{1},{3},{5},{7},{9},{11},{13},{15},{17},{19},{21},{23}},
xlabel style={font=\color{white!15!black}},
xlabel={Time (hour of the day)},
ymin=0,
ymax=1.5,
ylabel style={font=\color{white!15!black}},
ylabel={Nodal power (pu)},
axis background/.style={fill=white},
legend style={legend pos= north west, legend columns=3, legend cell align=left, align=left, draw=white!15!black}
]

\addplot[area legend, draw=none, fill=white!70!green, fill opacity=0.5]
table[row sep=crcr] {%
x	y\\
1	0.321111111111111\\
2	0.258716684692805\\
3	0.23628414051128\\
4	0.210676687547146\\
5	0.190341970120175\\
6	0.191556715016021\\
7	0.214684661412041\\
8	0.285012291623813\\
9	0.371947209873878\\
10	0.401081327184305\\
11	0.425990652195166\\
12	0.466234046605939\\
13	0.479641120937213\\
14	0.521898729998726\\
15	0.559611402218531\\
16	0.560102615959913\\
17	0.518786068948592\\
18	0.496400305739799\\
19	0.617697409936834\\
20	0.806217715559806\\
21	0.939346206086114\\
22	0.840005771092415\\
23	0.701698971267298\\
24	0.507137017538225\\
24	0.983937017538224\\
23	0.996098971267296\\
22	0.987205771092413\\
21	0.987885531928811\\
20	0.991817715559804\\
19	0.863647823159977\\
18	0.653200305739801\\
17	0.641576766623011\\
16	0.997749674783444\\
15	0.997258461042061\\
14	0.997650942388112\\
13	0.997871209432794\\
12	0.997207497933378\\
11	0.9994419796288\\
10	0.994022503654894\\
9	0.717993721501785\\
8	0.285012291623813\\
7	0.559808628354191\\
6	0.709309524004785\\
5	0.708741970120179\\
4	0.75147668754715\\
3	0.754036949500044\\
2	0.803435785816402\\
1	0.614664830119377\\
}--cycle;
\addlegendentry{Q0.00-Q1.00}

\addplot[area legend, draw=none, fill=white!40!green, fill opacity=0.5]
table[row sep=crcr] {%
x	y\\
1	0.321111111111111\\
2	0.258716684692805\\
3	0.23628414051128\\
4	0.210676687547146\\
5	0.190341970120175\\
6	0.191556715016021\\
7	0.214684661412041\\
8	0.285012291623813\\
9	0.371947209873878\\
10	0.401081327184305\\
11	0.425990652195166\\
12	0.466234046605939\\
13	0.479641120937213\\
14	0.521898729998726\\
15	0.559611402218531\\
16	0.560102615959913\\
17	0.518786068948592\\
18	0.496400305739799\\
19	0.617697409936834\\
20	0.806217715559806\\
21	0.939346206086114\\
22	0.840005771092415\\
23	0.701698971267298\\
24	0.507137017538225\\
24	0.791749672771814\\
23	0.862000551664574\\
22	0.905251530118078\\
21	0.94622339906857\\
20	0.909856979007682\\
19	0.795381620463149\\
18	0.544050008509119\\
17	0.577744944534226\\
16	0.959191367739698\\
15	0.991353215222806\\
14	0.983711734270206\\
13	0.993262383395685\\
12	0.982247335642487\\
11	0.98209886292132\\
10	0.921370292327889\\
9	0.478936554284209\\
8	0.285012291623813\\
7	0.416537292990989\\
6	0.521288482451138\\
5	0.504971519199484\\
4	0.506643049733604\\
3	0.527216429912468\\
2	0.575329704846495\\
1	0.431988304093567\\
}--cycle;
\addlegendentry{Q0.33-Q0.67}

\addplot [color=white!10!green, line width=2.0pt]
  table[row sep=crcr]{%
1	0.321111111111111\\
2	0.258716684692805\\
3	0.23628414051128\\
4	0.210676687547146\\
5	0.190341970120175\\
6	0.191556715016021\\
7	0.214684661412041\\
8	0.285012291623813\\
9	0.371947209873879\\
10	0.401081327184305\\
11	0.425990652195166\\
12	0.466234046605939\\
13	0.479641120937213\\
14	0.521898729998726\\
15	0.559611402218531\\
16	0.560102615959913\\
17	0.518786068948592\\
18	0.496400305739799\\
19	0.617697409936834\\
20	0.806217715559806\\
21	0.939346206086114\\
22	0.840005771092415\\
23	0.701698971267298\\
24	0.507137017538225\\
};
\addlegendentry{Q0.5}

\addplot[const plot, color=black, dashed, line width=2.0pt] table[row sep=crcr] {%
1	1\\
2	1\\
3	1\\
4	1\\
5	1\\
6	1\\
7	1\\
8	1\\
9	1\\
10	1\\
11	1\\
12	1\\
13	1\\
14	1\\
15	1\\
16	1\\
17	1\\
18	1\\
19	1\\
20	1\\
21	1\\
22	1\\
23	1\\
24	1\\
}; 

\end{axis}

\end{tikzpicture}%
} 
          \caption{Scenario A - SPC}
          \label{P_nodal_Scenario A - SPC}
     \end{subfigure}
          \begin{subfigure}[b]{0.5\textwidth}
          \centering
          \resizebox{\linewidth}{!}{
%
%
\begin{tikzpicture}

\begin{axis}[%
width=3.4in,
height=0.9in,
at={(0.758in,0.481in)},
scale only axis,
xmin=1,
xmax=24,
xtick={2,4,6,8,10,12,14,16,18,20,22,24},
xticklabels={{1},{3},{5},{7},{9},{11},{13},{15},{17},{19},{21},{23}},
xlabel style={font=\color{white!15!black}},
xlabel={Time (hour of the day)},
ymin=0,
ymax=1.5,
ylabel style={font=\color{white!15!black}},
ylabel={Nodal power (pu)},
axis background/.style={fill=white},
legend style={legend pos= north west, legend columns=3, legend cell align=left, align=left, draw=white!15!black}
]

\addplot[area legend, draw=none, fill=white!70!green, fill opacity=0.5]
table[row sep=crcr] {%
x	y\\
1	0.321111111111111\\
2	0.258716684692805\\
3	0.23628414051128\\
4	0.210676687547146\\
5	0.190341970120175\\
6	0.191556715016021\\
7	0.214684661412041\\
8	0.285012291623813\\
9	0.371947209873878\\
10	0.401081327184305\\
11	0.425990652195166\\
12	0.466234046605939\\
13	0.479641120937213\\
14	0.521898729998726\\
15	0.559611402218531\\
16	0.560102615959913\\
17	0.518786068948592\\
18	0.496400305739799\\
19	0.617697409936834\\
20	0.806217715559806\\
21	0.939346206086114\\
22	0.840005771092415\\
23	0.701698971267298\\
24	0.507137017538225\\
24	0.887963463819221\\
23	0.999298971267295\\
22	0.998683457042829\\
21	0.998850338317518\\
20	0.999901926086122\\
19	0.998523856217829\\
18	0.803600305739801\\
17	0.801139010125063\\
16	0.849514380665795\\
15	0.849023166924413\\
14	0.804251671175196\\
13	0.769052885643095\\
12	0.748586987782409\\
11	0.715402416901048\\
10	0.690493091890186\\
9	0.562535445167996\\
8	0.285012291623813\\
7	0.595511107693034\\
6	0.572383161297014\\
5	0.567941970120177\\
4	0.591503133828139\\
3	0.617110586792273\\
2	0.639543130973798\\
1	0.701911111111115\\
}--cycle;
\addlegendentry{Q0.00-Q1.00}

\addplot[area legend, draw=none, fill=white!40!green, fill opacity=0.5]
table[row sep=crcr] {%
x	y\\
1	0.321111111111111\\
2	0.258716684692805\\
3	0.23628414051128\\
4	0.210676687547146\\
5	0.190341970120175\\
6	0.191556715016021\\
7	0.214684661412041\\
8	0.285012291623813\\
9	0.371947209873878\\
10	0.401081327184305\\
11	0.425990652195166\\
12	0.466234046605939\\
13	0.479641120937213\\
14	0.521898729998726\\
15	0.559611402218531\\
16	0.560102615959913\\
17	0.518786068948592\\
18	0.496400305739799\\
19	0.617697409936834\\
20	0.806217715559806\\
21	0.939346206086114\\
22	0.840005771092415\\
23	0.701698971267298\\
24	0.507137017538225\\
24	0.871812751817744\\
23	0.996751208602209\\
22	0.992388976296436\\
21	0.997170767489622\\
20	0.995256623536408\\
19	0.982373144216353\\
18	0.721170495677453\\
17	0.551761863917743\\
16	0.593078410929064\\
15	0.601889522769076\\
14	0.562544197488047\\
13	0.511800589375751\\
12	0.515365513146043\\
11	0.476754771796494\\
10	0.423122143510835\\
9	0.410143697771363\\
8	0.285012291623813\\
7	0.586377939551209\\
6	0.55623244929554\\
5	0.528378797726381\\
4	0.575352421826665\\
3	0.588769849164936\\
2	0.623392418972323\\
1	0.685014257406602\\
}--cycle;
\addlegendentry{Q0.33-Q0.67}

\addplot [color=white!10!green, line width=2.0pt]
  table[row sep=crcr]{%
1	0.321111111111111\\
2	0.258716684692805\\
3	0.23628414051128\\
4	0.210676687547146\\
5	0.190341970120175\\
6	0.191556715016021\\
7	0.214684661412041\\
8	0.285012291623813\\
9	0.371947209873879\\
10	0.401081327184305\\
11	0.425990652195166\\
12	0.466234046605939\\
13	0.479641120937213\\
14	0.521898729998726\\
15	0.559611402218531\\
16	0.560102615959913\\
17	0.518786068948592\\
18	0.496400305739799\\
19	0.617697409936834\\
20	0.806217715559806\\
21	0.939346206086114\\
22	0.840005771092415\\
23	0.701698971267298\\
24	0.507137017538225\\
};
\addlegendentry{Q0.5}

\addplot[const plot, color=black, dashed, line width=2.0pt] table[row sep=crcr] {%
1	1\\
2	1\\
3	1\\
4	1\\
5	1\\
6	1\\
7	1\\
8	1\\
9	1\\
10	1\\
11	1\\
12	1\\
13	1\\
14	1\\
15	1\\
16	1\\
17	1\\
18	1\\
19	1\\
20	1\\
21	1\\
22	1\\
23	1\\
24	1\\
}; 

\end{axis}

\end{tikzpicture}%
}
          \caption{Scenario A - MPC}
          \label{P_nodal_Scenario A - MPC}
     \end{subfigure}\\
     \begin{subfigure}[b]{0.5\textwidth}
          \centering
          \resizebox{\linewidth}{!}{
%
%
\begin{tikzpicture}

\begin{axis}[%
width=3.4in,
height=0.9in,
at={(0.758in,0.481in)},
scale only axis,
xmin=1,
xmax=24,
xtick={2,4,6,8,10,12,14,16,18,20,22,24},
xticklabels={{1},{3},{5},{7},{9},{11},{13},{15},{17},{19},{21},{23}},
xlabel style={font=\color{white!15!black}},
xlabel={Time (hour of the day)},
ymin=0,
ymax=1.5,
ylabel style={font=\color{white!15!black}},
ylabel={Nodal power (pu)},
axis background/.style={fill=white},
legend style={legend pos= north west, legend columns=3, legend cell align=left, align=left, draw=white!15!black}
]

\addplot[area legend, draw=none, fill=white!70!green, fill opacity=0.5]
table[row sep=crcr] {%
x	y\\
1	0.321111111111111\\
2	0.258716684692805\\
3	0.23628414051128\\
4	0.210676687547146\\
5	0.190341970120175\\
6	0.191556715016021\\
7	0.214684661412041\\
8	0.285012291623813\\
9	0.371947209873878\\
10	0.401081327184305\\
11	0.425990652195166\\
12	0.466234046605939\\
13	0.479641120937213\\
14	0.521898729998726\\
15	0.559611402218531\\
16	0.560102615959913\\
17	0.518786068948592\\
18	0.496400305739799\\
19	0.617697409936834\\
20	0.806217715559806\\
21	0.939346206086114\\
22	0.840005771092415\\
23	0.701698971267298\\
24	0.507137017538225\\
24	0.958337017538225\\
23	0.919298971267297\\
22	0.958405771092413\\
21	0.996946206086113\\
20	0.959817715559804\\
19	0.812897409936835\\
18	0.538000305739799\\
17	0.850550774830945\\
16	0.997749674783444\\
15	0.997258461042061\\
14	0.994839906469315\\
13	0.993129493030235\\
12	0.979722418698961\\
11	0.939479024288188\\
10	0.914569699277327\\
9	0.851947209873878\\
8	0.285012291623813\\
7	0.601884661412044\\
6	0.719556715016025\\
5	0.651141970120178\\
4	0.68107668754715\\
3	0.578684140511282\\
2	0.517916684692806\\
1	0.436311111111112\\
}--cycle;
\addlegendentry{Q0.00-Q1.00}

\addplot[area legend, draw=none, fill=white!40!green, fill opacity=0.5]
table[row sep=crcr] {%
x	y\\
1	0.321111111111111\\
2	0.258716684692805\\
3	0.23628414051128\\
4	0.210676687547146\\
5	0.190341970120175\\
6	0.191556715016021\\
7	0.214684661412041\\
8	0.285012291623813\\
9	0.371947209873878\\
10	0.401081327184305\\
11	0.425990652195166\\
12	0.466234046605939\\
13	0.479641120937213\\
14	0.521898729998726\\
15	0.559611402218531\\
16	0.560102615959913\\
17	0.518786068948592\\
18	0.496400305739799\\
19	0.617697409936834\\
20	0.806217715559806\\
21	0.939346206086114\\
22	0.840005771092415\\
23	0.701698971267298\\
24	0.507137017538225\\
24	0.620458523550446\\
23	0.779912687375752\\
22	0.854387793564325\\
21	0.949060379159246\\
20	0.843138115733795\\
19	0.671958682954088\\
18	0.501028404913353\\
17	0.786939842085754\\
16	0.936147405557172\\
15	0.937613930663541\\
14	0.89136233845096\\
13	0.955559488284156\\
12	0.899949054733088\\
11	0.794038331443859\\
10	0.790953098901844\\
9	0.661237439239962\\
8	0.285012291623813\\
7	0.379100194627101\\
6	0.371681406911048\\
5	0.43296339972725\\
4	0.398480080326615\\
3	0.381823269231957\\
2	0.361251673851093\\
1	0.370580852881267\\
}--cycle;
\addlegendentry{Q0.33-Q0.67}

\addplot [color=white!10!green, line width=2.0pt]
  table[row sep=crcr]{%
1	0.321111111111111\\
2	0.258716684692805\\
3	0.23628414051128\\
4	0.210676687547146\\
5	0.190341970120175\\
6	0.191556715016021\\
7	0.214684661412041\\
8	0.285012291623813\\
9	0.371947209873879\\
10	0.401081327184305\\
11	0.425990652195166\\
12	0.466234046605939\\
13	0.479641120937213\\
14	0.521898729998726\\
15	0.559611402218531\\
16	0.560102615959913\\
17	0.518786068948592\\
18	0.496400305739799\\
19	0.617697409936834\\
20	0.806217715559806\\
21	0.939346206086114\\
22	0.840005771092415\\
23	0.701698971267298\\
24	0.507137017538225\\
};
\addlegendentry{Q0.5}

\addplot[const plot, color=black, dashed, line width=2.0pt] table[row sep=crcr] {%
1	1\\
2	1\\
3	1\\
4	1\\
5	1\\
6	1\\
7	1\\
8	1\\
9	1\\
10	1\\
11	1\\
12	1\\
13	1\\
14	1\\
15	1\\
16	1\\
17	1\\
18	1\\
19	1\\
20	1\\
21	1\\
22	1\\
23	1\\
24	1\\
}; 

\end{axis}

\end{tikzpicture}%
}
          \caption{Scenario B - SPC}
          \label{P_nodal_Scenario B - SPC}
     \end{subfigure}
     \begin{subfigure}[b]{0.5\textwidth}
          \centering
          \resizebox{\linewidth}{!}{
%
%
\begin{tikzpicture}

\begin{axis}[%
width=3.4in,
height=0.9in,
at={(0.758in,0.481in)},
scale only axis,
xmin=1,
xmax=24,
xtick={2,4,6,8,10,12,14,16,18,20,22,24},
xticklabels={{1},{3},{5},{7},{9},{11},{13},{15},{17},{19},{21},{23}},
xlabel style={font=\color{white!15!black}},
xlabel={Time (hour of the day)},
ymin=0,
ymax=1.5,
ylabel style={font=\color{white!15!black}},
ylabel={Nodal power (pu)},
axis background/.style={fill=white},
legend style={legend pos= north west, legend columns=3, legend cell align=left, align=left, draw=white!15!black}
]

\addplot[area legend, draw=none, fill=white!70!green, fill opacity=0.5]
table[row sep=crcr] {%
x	y\\
1	0.321111111111111\\
2	0.258716684692805\\
3	0.23628414051128\\
4	0.210676687547146\\
5	0.190341970120175\\
6	0.191556715016021\\
7	0.214684661412041\\
8	0.285012291623813\\
9	0.371947209873878\\
10	0.401081327184305\\
11	0.425990652195166\\
12	0.466234046605939\\
13	0.479641120937213\\
14	0.521898729998726\\
15	0.559611402218531\\
16	0.560102615959913\\
17	0.518786068948592\\
18	0.496400305739799\\
19	0.617697409936834\\
20	0.806217715559806\\
21	0.939346206086114\\
22	0.840005771092415\\
23	0.701698971267298\\
24	0.507137017538225\\
24	0.791937017538227\\
23	0.999219632424325\\
22	0.998683457042829\\
21	0.998850338317518\\
20	0.999901926086122\\
19	0.902497409936833\\
18	0.673242411002957\\
17	0.90702136306624\\
16	0.955396733606973\\
15	0.95490551986559\\
14	0.910134024116373\\
13	0.874935238584272\\
12	0.861528164252997\\
11	0.821284769842224\\
10	0.796375444831363\\
9	0.760182503991525\\
8	0.285012291623813\\
7	0.501000450885725\\
6	0.494714609752863\\
5	0.493499864857017\\
4	0.508197348704171\\
3	0.533804801668305\\
2	0.561874579429646\\
1	0.618631772268137\\
}--cycle;
\addlegendentry{Q0.00-Q1.00}

\addplot[area legend, draw=none, fill=white!40!green, fill opacity=0.5]
table[row sep=crcr] {%
x	y\\
1	0.321111111111111\\
2	0.258716684692805\\
3	0.23628414051128\\
4	0.210676687547146\\
5	0.190341970120175\\
6	0.191556715016021\\
7	0.214684661412041\\
8	0.285012291623813\\
9	0.371947209873878\\
10	0.401081327184305\\
11	0.425990652195166\\
12	0.466234046605939\\
13	0.479641120937213\\
14	0.521898729998726\\
15	0.559611402218531\\
16	0.560102615959913\\
17	0.518786068948592\\
18	0.496400305739799\\
19	0.617697409936834\\
20	0.806217715559806\\
21	0.939346206086114\\
22	0.840005771092415\\
23	0.701698971267298\\
24	0.507137017538225\\
24	0.739693690118553\\
23	0.964186651141141\\
22	0.992388976296436\\
21	0.997170767489622\\
20	0.995256623536408\\
19	0.820275765939199\\
18	0.623518763040076\\
17	0.662744303405166\\
16	0.705693503477711\\
15	0.705202289736328\\
14	0.694580267730098\\
13	0.625232008455011\\
12	0.629613258755915\\
11	0.597039536865313\\
10	0.555974540283498\\
9	0.536142748554467\\
8	0.285012291623813\\
7	0.481116084942545\\
6	0.46320914947295\\
5	0.450161823402745\\
4	0.45912927970169\\
3	0.508469908301542\\
2	0.530902452483067\\
1	0.569563703265655\\
}--cycle;
\addlegendentry{Q0.33-Q0.67}

\addplot [color=white!10!green, line width=2.0pt]
  table[row sep=crcr]{%
1	0.321111111111111\\
2	0.258716684692805\\
3	0.23628414051128\\
4	0.210676687547146\\
5	0.190341970120175\\
6	0.191556715016021\\
7	0.214684661412041\\
8	0.285012291623813\\
9	0.371947209873879\\
10	0.401081327184305\\
11	0.425990652195166\\
12	0.466234046605939\\
13	0.479641120937213\\
14	0.521898729998726\\
15	0.559611402218531\\
16	0.560102615959913\\
17	0.518786068948592\\
18	0.496400305739799\\
19	0.617697409936834\\
20	0.806217715559806\\
21	0.939346206086114\\
22	0.840005771092415\\
23	0.701698971267298\\
24	0.507137017538225\\
};
\addlegendentry{Q0.5}

\addplot[const plot, color=black, dashed, line width=2.0pt] table[row sep=crcr] {%
1	1\\
2	1\\
3	1\\
4	1\\
5	1\\
6	1\\
7	1\\
8	1\\
9	1\\
10	1\\
11	1\\
12	1\\
13	1\\
14	1\\
15	1\\
16	1\\
17	1\\
18	1\\
19	1\\
20	1\\
21	1\\
22	1\\
23	1\\
24	1\\
}; 

\end{axis}

\end{tikzpicture}%
}
          \caption{Scenario B - MPC}
          \label{P_nodal_Scenario B - MPC}
     \end{subfigure}
    \caption{Active power flow in the substation transformers located at the multiple nodes of CIGRE MV grid, over the charging horizon (base case for all scenarios). The green shaded bands denote different quantile intervals across all the 14-nodes.}\label{P_nodal}
\end{figure*}


Fig. \ref{P_nodal} shows the active nodal injections (conventional demand + EVs): nodal injections are re-scaled by the rated power of each node, so that 1 per unit (dashed lines) corresponds to the maximum power flow at that node; the shaded bands refer to different quantiles, whereas the thicker line is the average value. It is possible to observe in Fig.~\ref{P_nodal} that, in the evening, the power flows hit the limit in all the cases. This is due to the confluence of evening conventional demand and the one of the EVs. With SPCs, the grid is overloaded in the day's central part also, whereas less with MPCs. This denotes that MPCs tend to shift the charging demand during the afternoon and evening hours. This is also in line with the previous findings, where the MPCs case features more chargers in the nodes corresponding to the overnight parking locations.

\subsection{Sensitivity of the results to the charging demand}
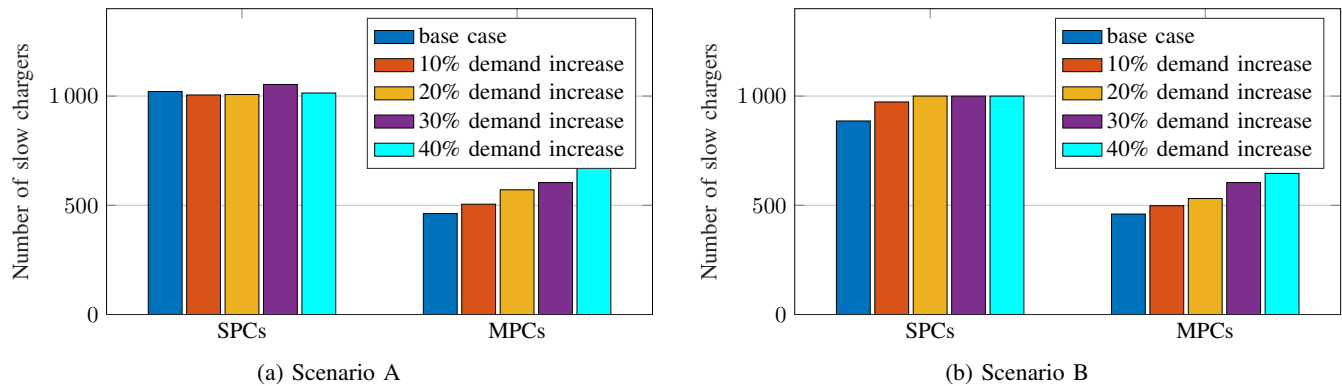
\begin{figure*}[!h]
     \begin{subfigure}[b]{0.5\textwidth}
          \centering
          \resizebox{0.95\textwidth}{!}{
%
%
\definecolor{mycolor1}{rgb}{0.00000,0.44700,0.74100}%
\definecolor{mycolor2}{rgb}{0.85000,0.32500,0.09800}%
\definecolor{mycolor3}{rgb}{0.92900,0.69400,0.12500}%
\definecolor{mycolor4}{rgb}{0.49400,0.18400,0.55600}%
\definecolor{mycolor5}{rgb}{0.00000,1.00000,1.00000}%
\begin{tikzpicture}

\begin{axis}[%
width=3.3in,
height=1.85in,
at={(0.759in,0.481in)},
scale only axis,
bar shift auto,
xmin=0.507692307692308,
xmax=2.49230769230769,
xtick={1,2},
xticklabels={{SPCs},{MPCs}},
ymin=0,
ymax=1400,
ylabel style={font=\color{white!15!black}},
ylabel={Number of slow chargers},
axis background/.style={fill=white},
ymajorgrids,
legend style={legend pos= north east, legend columns=1, legend cell align=left, align=left, draw=white!15!black}
]
\addplot[ybar, bar width=0.123, fill=mycolor1, draw=black, area legend] table[row sep=crcr] {%
1	1021\\
2	462\\
};
\addplot[forget plot, color=white!15!black] table[row sep=crcr] {%
0.507692307692308	0\\
2.49230769230769	0\\
};
\addlegendentry{base case}

\addplot[ybar, bar width=0.123, fill=mycolor2, draw=black, area legend] table[row sep=crcr] {%
1	1005\\
2	505\\
};
\addplot[forget plot, color=white!15!black] table[row sep=crcr] {%
0.507692307692308	0\\
2.49230769230769	0\\
};
\addlegendentry{10\% demand increase}

\addplot[ybar, bar width=0.123, fill=mycolor3, draw=black, area legend] table[row sep=crcr] {%
1	1007\\
2	571\\
};
\addplot[forget plot, color=white!15!black] table[row sep=crcr] {%
0.507692307692308	0\\
2.49230769230769	0\\
};
\addlegendentry{20\% demand increase}

\addplot[ybar, bar width=0.123, fill=mycolor4, draw=black, area legend] table[row sep=crcr] {%
1	1053\\
2	604\\
};
\addplot[forget plot, color=white!15!black] table[row sep=crcr] {%
0.507692307692308	0\\
2.49230769230769	0\\
};
\addlegendentry{30\% demand increase}

\addplot[ybar, bar width=0.123, fill=mycolor5, draw=black, area legend] table[row sep=crcr] {%
1	1014\\
2	686\\
};
\addplot[forget plot, color=white!15!black] table[row sep=crcr] {%
0.507692307692308	0\\
2.49230769230769	0\\
};
\addlegendentry{40\% demand increase}

\end{axis}

\end{tikzpicture}%
}
          \caption{Scenario A}
          \label{SensitivityPlots_ScenarioA}
     \end{subfigure}
     \begin{subfigure}[b]{0.5\textwidth}
          \centering
          \resizebox{0.95\textwidth}{!}{
%
%
\definecolor{mycolor1}{rgb}{0.00000,0.44700,0.74100}%
\definecolor{mycolor2}{rgb}{0.85000,0.32500,0.09800}%
\definecolor{mycolor3}{rgb}{0.92900,0.69400,0.12500}%
\definecolor{mycolor4}{rgb}{0.49400,0.18400,0.55600}%
\definecolor{mycolor5}{rgb}{0.00000,1.00000,1.00000}%
\begin{tikzpicture}

\begin{axis}[%
width=3.3in,
height=1.85in,
at={(0.759in,0.481in)},
scale only axis,
bar shift auto,
xmin=0.507692307692308,
xmax=2.49230769230769,
xtick={1,2},
xticklabels={{SPCs},{MPCs}},
ymin=0,
ymax=1400,
ylabel style={font=\color{white!15!black}},
ylabel={Number of slow chargers},
axis background/.style={fill=white},
ymajorgrids,
legend style={legend pos= north east, legend columns=1, legend cell align=left, align=left, draw=white!15!black}
]
\addplot[ybar, bar width=0.123, fill=mycolor1, draw=black, area legend] table[row sep=crcr] {%
1	886\\
2	460\\
};
\addplot[forget plot, color=white!15!black] table[row sep=crcr] {%
0.507692307692308	0\\
2.49230769230769	0\\
};
\addlegendentry{base case}

\addplot[ybar, bar width=0.123, fill=mycolor2, draw=black, area legend] table[row sep=crcr] {%
1	973\\
2	498\\
};
\addplot[forget plot, color=white!15!black] table[row sep=crcr] {%
0.507692307692308	0\\
2.49230769230769	0\\
};
\addlegendentry{10\% demand increase}

\addplot[ybar, bar width=0.123, fill=mycolor3, draw=black, area legend] table[row sep=crcr] {%
1	1000\\
2	531\\
};
\addplot[forget plot, color=white!15!black] table[row sep=crcr] {%
0.507692307692308	0\\
2.49230769230769	0\\
};
\addlegendentry{20\% demand increase}

\addplot[ybar, bar width=0.123, fill=mycolor4, draw=black, area legend] table[row sep=crcr] {%
1	1000\\
2	604\\
};
\addplot[forget plot, color=white!15!black] table[row sep=crcr] {%
0.507692307692308	0\\
2.49230769230769	0\\
};
\addlegendentry{30\% demand increase}

\addplot[ybar, bar width=0.123, fill=mycolor5, draw=black, area legend] table[row sep=crcr] {%
1	1000\\
2	646\\
};
\addplot[forget plot, color=white!15!black] table[row sep=crcr] {%
0.507692307692308	0\\
2.49230769230769	0\\
};
\addlegendentry{40\% demand increase}

\end{axis}

\end{tikzpicture}%
} 
          \caption{Scenario B}
          \label{SensitivityPlots_ScenarioB}
     \end{subfigure}
     \caption{Sensitivity analysis with increased energy demand by EVs for different scenarios.}\label{SensitivityPlots}
\end{figure*}

We analyze the number of required slow chargers for increasing values of the driving demand (hence, of the charging demand), from +10\% to +40\% of the case considered so far. This analysis is shown in Fig. \ref{SensitivityPlots}.

We preliminary point out that the decrease in the number of chargers observable for Scenario A/SPCs is not justifiable reasonably since we expect it to be non-decreasing for higher charging needs. In fact, the increasing number of chargers is due to the MIP gap setting used to solve the optimization problem (10\%), which ultimately results in a differently approximated problem solution. In other words, the small variations in the number of chargers in Scenario A/SPC and small decreasing and increasing trends are because of approximated solutions and are not of particular significance. The relevant conclusions inferred from Fig. \ref{SensitivityPlots} are the following.

\textbf{Finding 6}: the SPCs solution is not significantly impacted by lower or higher charging demand. At all demand levels, approximately 1'000 chargers (i.e., around 1 per EV) are sufficient to cover the demand. Increased charging needs are absorbed by the spare capacity of chargers, without requiring new ones. The SPCs solution, although more expensive, is more robust against changes of the driving demand.

\textbf{Finding 7}: the MPCs solution is more sensitive to increasing levels of demand than SPCs. Symmetrically to the former finding, optimized utilization of the MPCs saturates their capacity, requiring new chargers for increased charging needs.


\subsection{Computational performance of the algorithm}
The optimization problem is implemented in MATLAB and solved using Gurobi. Solving the problem for 1'000 EVs required approximately 90 minutes with a MIP gap of 10\%. The MIP gap setting is chosen as a trade-off between computational performance and accuracy of the solution.

\section{Conclusions}
This paper presented a method for cost-optimal planning of the charging infrastructure of EVs. The formulation considered both slow, fast, single-port, and multi-port chargers. In addition, it included the availability of the drivers to plug and unplug their vehicles for optimized utilization of the charging infrastructure and power grid constraints to model voltage limits, current limits, and rating of the substation transformer. By suitably modifying nonlinear constraints appearing in the formulation, we derived a mixed-integer linear formulation of the problem that could be solved with off-the-shelf software libraries and in a reasonable time. The method was applied on a 14-bus MV network considering a population of 1'000 EVs. The most notable indication from the results is that MPCs achieve the lowest infrastructure cost compared to other options, including increased drivers' flexibility, ultimately denoting that this technological solution can substantially improve the charge arbitrage independently of possibly hard-to-predict consumer behavior.


\bibliographystyle{IEEEtran}
\bibliography{./sample}

\begin{thebibliography}{10}
\providecommand{\url}[1]{#1}
\csname url@samestyle\endcsname
\providecommand{\newblock}{\relax}
\providecommand{\bibinfo}[2]{#2}
\providecommand{\BIBentrySTDinterwordspacing}{\spaceskip=0pt\relax}
\providecommand{\BIBentryALTinterwordstretchfactor}{4}
\providecommand{\BIBentryALTinterwordspacing}{\spaceskip=\fontdimen2\font plus
\BIBentryALTinterwordstretchfactor\fontdimen3\font minus
  \fontdimen4\font\relax}
\providecommand{\BIBforeignlanguage}[2]{{%
\expandafter\ifx\csname l@#1\endcsname\relax
\typeout{** WARNING: IEEEtran.bst: No hyphenation pattern has been}%
\typeout{** loaded for the language `#1'. Using the pattern for}%
\typeout{** the default language instead.}%
\else
\language=\csname l@#1\endcsname
\fi
#2}}
\providecommand{\BIBdecl}{\relax}
\BIBdecl

\bibitem{EUCEF}
\BIBentryALTinterwordspacing
``{EU} 2030 climate \& energy framework,'' accessed: 2020-07. [Online].
  Available:
  \url{https://www.consilium.europa.eu/en/policies/climate-change/2030-climate-and-energy-framework/}
\BIBentrySTDinterwordspacing

\bibitem{1irena_smartcharging}
{IRENA}, ``Innovation outlook: Smart charging for electric vehicles,''
  International Renewable Energy Agency, Abu Dhabi, Tech. Rep. MSU-CSE-06-2,
  2019.

\bibitem{metais:hal-03127266}
\BIBentryALTinterwordspacing
M.-O. Metais, O.~Jouini, Y.~Perez, J.~Berrada, and E.~Suomalainen, ``{Too much
  or not enough? Planning electric vehicle charging infrastructure: a review of
  modeling options.}'' Feb. 2021, working paper or preprint. [Online].
  Available: \url{https://hal.archives-ouvertes.fr/hal-03127266}
\BIBentrySTDinterwordspacing

\bibitem{EV_dataPlot}
``{IEA}, {E}lectric car stock by region and technology, 2013-2018, {IEA},
  {P}aris,''
  \url{https://www.iea.org/data-and-statistics/charts/electric-car-stock-by-region-and-technology-2013-2018},
  accessed: 2021-08.

\bibitem{USAInvestment}
\BIBentryALTinterwordspacing
``Estimating electric vehicle charging infrastructure costs across major {U.S.}
  metropolitan areas,'' accessed: 2021-03. [Online]. Available:
  \url{https://theicct.org/publications/charging-cost-US}
\BIBentrySTDinterwordspacing

\bibitem{MinistryReport}
\BIBentryALTinterwordspacing
``Ministère de l’environnement de l’énergie et de lamer - {L}a loi de
  transition Énergétique pour la croissance verte,'' accessed: 2021-03.
  [Online]. Available: \url{https://bit.ly/3aRJ1Lb}
\BIBentrySTDinterwordspacing

\bibitem{6674071}
M.~S. ElNozahy and M.~M.~A. Salama, ``A comprehensive study of the impacts of
  {PHEV}s on residential distribution networks,'' \emph{IEEE Transactions on
  Sustainable Energy}, vol.~5, no.~1, pp. 332--342, 2014.

\bibitem{en12244717}
S.~Johansson, J.~Persson, S.~Lazarou, and A.~Theocharis, ``Investigation of the
  impact of large-scale integration of electric vehicles for a swedish
  distribution network,'' \emph{Energies}, vol.~12, 2019.

\bibitem{HengsongWang}
H.~Wang, Q.~Huang, C.~Zhang, and A.~Xia, ``\BIBforeignlanguage{Undetermined}{A
  novel approach for the layout of electric vehicle charging station}.''

\bibitem{6398568}
Z.~Hu and Y.~Song, ``Distribution network expansion planning with optimal
  siting and sizing of electric vehicle charging stations,'' in \emph{2012 47th
  International Universities Power Engineering Conference (UPEC)}, 2012, pp.
  1--6.

\bibitem{6362255}
Z.~Liu, F.~Wen, and G.~Ledwich, ``Optimal planning of electric-vehicle charging
  stations in distribution systems,'' \emph{IEEE Transactions on Power
  Delivery}, vol.~28, no.~1, pp. 102--110, 2013.

\bibitem{6879337}
A.~Y.~S. Lam, Y.-W. Leung, and X.~Chu, ``Electric vehicle charging station
  placement: Formulation, complexity, and solutions,'' \emph{IEEE Transactions
  on Smart Grid}, vol.~5, no.~6, pp. 2846--2856, 2014.

\bibitem{7368203}
C.~Luo, Y.-F. Huang, and V.~Gupta, ``Placement of {EV} charging
  stations—balancing benefits among multiple entities,'' \emph{IEEE
  Transactions on Smart Grid}, vol.~8, no.~2, pp. 759--768, 2017.

\bibitem{8000654}
H.~Chen, Z.~Hu, H.~Luo, J.~Qin, R.~Rajagopal, and H.~Zhang, ``Design and
  planning of a multiple-charger multiple-port charging system for {PEV}
  charging station,'' \emph{IEEE Transactions on Smart Grid}, vol.~10, no.~1,
  pp. 173--183, 2019.

\bibitem{en12132595}
G.~Liu, L.~Kang, Z.~Luan, J.~Qiu, and F.~Zheng, ``Charging station and power
  network planning for integrated electric vehicles ({EV}s),'' \emph{Energies},
  vol.~12, no.~13, 2019.

\bibitem{XIANG2016647}
Y.~Xiang, J.~Liu, R.~Li, F.~Li, C.~Gu, and S.~Tang, ``Economic planning of
  electric vehicle charging stations considering traffic constraints and load
  profile templates,'' \emph{Applied Energy}, vol. 178, pp. 647--659, 2016.

\bibitem{8366991}
J.~{Li}, X.~{Sun}, Q.~{Liu}, W.~{Zheng}, H.~{Liu}, and J.~A. {Stankovic},
  ``Planning electric vehicle charging stations based on user charging
  behavior,'' in \emph{2018 IEEE/ACM Third International Conference on
  Internet-of-Things Design and Implementation (IoTDI)}, 2018, pp. 225--236.

\bibitem{9265279}
W.~Gan, M.~Shahidehpour, J.~Guo, W.~Yao, A.~Paaso, L.~Zhang, and J.~Wen,
  ``Two-stage planning of network-constrained hybrid energy supply stations for
  electric and natural gas vehicles,'' \emph{IEEE Transactions on Smart Grid},
  vol.~12, no.~3, pp. 2013--2026, 2021.

\bibitem{9076708}
W.~Gan, M.~Shahidehpour, M.~Yan, J.~Guo, W.~Yao, A.~Paaso, L.~Zhang, and
  J.~Wen, ``Coordinated planning of transportation and electric power networks
  with the proliferation of electric vehicles,'' \emph{IEEE Transactions on
  Smart Grid}, vol.~11, no.~5, pp. 4005--4016, 2020.

\bibitem{ISGT2021V2G}
B.~{Mukherjee}, G.~{Kariniotakis}, and F.~{Sossan}, ``Smart charging,
  vehicle-to-grid, and reactive power support from electric vehicles in
  distribution grids: A performance comparison,'' in \emph{2021 IEEE PES ISGT
  Finland (ISGT Europe)}, 2021.

\bibitem{JuanVanRoyThesis}
J.~{Van Roy}, ``Electric vehicle charging integration in buildings: Local
  charging coordination and dc grids,'' Ph.D. dissertation, KU Leuven, 2015.

\bibitem{de2013plug}
L.~De~Vroey, R.~Jahn, M.~El~Baghdadi, and J.~Van~Mierlo, ``Plug-to-wheel energy
  balance-results of a two years experience behind the wheel of electric
  vehicles,'' \emph{World Electric Vehicle Journal}, vol.~6, no.~1, pp.
  130--134, 2013.

\bibitem{6473866}
K.~{Christakou}, J.~{LeBoudec}, M.~{Paolone}, and D.~{Tomozei}, ``Efficient
  computation of sensitivity coefficients of node voltages and line currents in
  unbalanced radial electrical distribution networks,'' \emph{IEEE Transactions
  on Smart Grid}, vol.~4, no.~2, pp. 741--750, 2013.

\bibitem{9243122}
F.~{Sossan}, B.~{Mukherjee}, and Z.~{Hu}, ``Impact of the charging demand of
  electric vehicles on distribution grids: a comparison between autonomous and
  non-autonomous driving,'' in \emph{2020 Fifteenth International Conference on
  Ecological Vehicles and Renewable Energies (EVER)}, 2020, pp. 1--6.

\bibitem{CIRED2021}
B.~Mukherjee, G.~Kariniotakis, and F.~Sossan, ``Scheduling the charge of
  electric vehicles including reactive power support: Application to a medium
  voltage grid,'' in \emph{CIRED}, 2021.

\bibitem{nick2014optimal}
M.~Nick, R.~Cherkaoui, and M.~Paolone, ``Optimal allocation of dispersed energy
  storage systems in active distribution networks for energy balance and grid
  support,'' \emph{IEEE Transactions on Power Systems}, vol.~29, no.~5, 2014.

\bibitem{en8031760}
N.~Leemput, F.~Geth, J.~Van~Roy, P.~Olivella-Rosell, J.~Driesen, and A.~Sumper,
  ``{MV} and {LV} residential grid impact of combined slow and fast charging of
  electric vehicles,'' \emph{Energies}, vol.~8, no.~3, 2015.

\bibitem{testanevrepo}
``Test-an-{EV} project: Electrical vehicle ({EV}) data,''
  \url{http://mclabprojects.di.uniroma1.it/smarthgnew/Test-an-EV/?EV-code=EV8},
  accessed: 2020-02.

\bibitem{PriceEUreport}
{Spöttle, M., Jörling, K., Schimmel, M., Staats, M., Grizzel L., Jerram, L.,
  Drier, W., Gartner, J.}, ``Research for {TRAN} committee - charging
  infrastructure for electric road vehicles,'' European Parliament, Policy
  Department for Structural and Cohesion Policies, Brussels, Tech. Rep., 2018.

\bibitem{RMIreport}
{Chris Nelder and Emily Rogers}, ``Reducing {EV} charging infrastructure
  costs,'' Rocky Mountain Institute, Tech. Rep., 2019.

\bibitem{slowChargerPrice}
\BIBentryALTinterwordspacing
``Schneider {EVlink} {G4} smart charging station - {EVB1A22P4ERI},'' accessed:
  2021-08. [Online]. Available: \url{https://bit.ly/3AVUWSO}
\BIBentrySTDinterwordspacing

\bibitem{CIGREREF}
{{CIGRE' Task Force C6.04.02}}, ``Benchmark systems for network integration of
  renewable and distributed energy resources,'' {CIGRE} International Council
  on large electric systems, Tech. Rep., July 2009.

\end{thebibliography}

\clearpage

\end{document}